\documentclass{iopart}[12pt]
\usepackage{iopams}
\eqnobysec
\def\r{\mathbb R}                   
\def\NAT{\mathbb N}                                       
\def\d{\partial}
\def\nb{\nabla}
\def\fr{\frac}
\def\be{\begin{equation}}
\def\ee{\end{equation}}
\def\bea{\begin{eqnarray}}
\def\eea{\end{eqnarray}}
\def\bnr{\begin{eqnarray*}}
\def\enr{\end{eqnarray*}}
\def\N{\hfill \raisebox{1mm}{\framebox{\rule{0mm}{1mm}}}}
\def\L{{\mathbb L}}
\def\T{{\bf T}}
\def\S{{\bf S}}
\def\p{{\bf P}}
\def\Q{{\bf Q}}
\def\G{{\bf g}}
\def\R{{\bf R}}
\def\U{{\bf U}}
\def\tT{\tilde{\G}}
\def\V{\vec{\bf v}}
\def\rmg{{\rm g}}
\def\DP{{\cal DP}}

\def\C{{\cal C}}

\def\g{\gamma}
\def\Z{\Theta}
\def\z{\btheta}
\def\s{\sigma}
\def\O{{\bf\Omega}}
\def\Sg{{\bf\Sigma}} 
\def\f{\varphi}

\def\P{{\bf Proof :} \hspace{3mm}}
\def\k{\vec{\bf k}}

\def\Nm{{\cal N}}
\def\Um{{\cal U}}

\def\l{\lambda}
\def\xiv{\vec{\bxi}}
\def\lie{{\pounds}_{\xiv}\, }

\def\vu{\mathbf u}
\def\n{\vec{\mathbf n}}
\def\u{\vec{\mathbf u}}
\def\v{\vec{\mathbf v}}
\def\K{{\mathbf k}}

\def\Diff{{\it Diff}}
\def\Conf{{\it Conf}} 
\newtheorem{defi}{Definition}[section]
\newtheorem{theo}{Theorem}[section]
\newtheorem{coro}{Corollary}[section]
\newtheorem{prop}{Proposition}[section]
\newtheorem{lem}{Lemma}[section]

\begin{document}
\title[General study and basic properties of causal symmetries]
{General study and basic properties of causal symmetries}

\author{Alfonso Garc\'{\i}a-Parrado and Jos\'{e} M M Senovilla}

\address{Departamento de F\'{\i}sica Te\'{o}rica,
 Universidad del Pa\'{\i}s Vasco, Apartado 644, 48080 Bilbao, Spain}

\eads{\mailto{wtbgagoa@lg.ehu.es} and \mailto{wtpmasej@lg.ehu.es}}

\begin{abstract}
We fully develop the concept of causal symmetry introduced in \cite{LETTER}. 
A causal symmetry is a transformation of a Lorentzian manifold $(V,\G)$ 
which maps every future-directed vector onto a future-directed vector. 

We prove that the set of all causal symmetries
is not a group under the usual composition operation 
but a {\it submonoid} of the diffeomorphism group of $V$. Therefore, 
the infinitesimal generating vector fields of 
causal symmetries ---{\em causal-preserving vector fields}--- are associated to 
local one-parameter groups of transformations
which are causal symmetries only for positive values of the 
parameter ---{\em one-parameter submonoids of causal symmetries}---. 
The pull-back of the metric
under each causal symmetry results in a new rank-2 future
tensor, and we prove that there is always a set of null directions canonical 
to the causal symmetry. As a result of this it makes sense to classify causal 
symmetries according to the number of independent
canonical null directions. This classification 
is maintained at the infinitesimal level where we find the 
necessary and sufficient conditions for a vector field to be causal 
preserving. 
They involve the Lie derivatives of the metric tensor 
and of the canonical null directions. 
In addition, we prove a stability property of these equations
under the repeated application of the Lie operator. Monotonicity 
properties, constants of motion and conserved currents
can be defined or built using casual preserving vector fields. Some 
illustrative examples are presented. 
\end{abstract}
\pacs{04.20.Cv,04.20.Gz,02.40.-k,02.20.-a}

\section{Introduction}
Symmetry is one of the truly important concepts in Physics. For 
gravitational theories, such as General Relativity and the like, 
symmetries have always been taken into account in many fruitful 
ways, in particular in order to analyse and understand the 
solutions to Einstein's field equations \cite{KRAMERS} with groups of 
motions (infinitesimal isometries). This analysis has been partly
generalized considering more general symmetries, such as homothetic or 
conformal Killing vectors. Nevertheless, as far as we know, the question of 
whether there are vector fields which, in some sense, preserve the causal 
structure of the spacetime has not been considered in full generality. 
This is our aim in this paper. To that end, we will base our study on the 
concept of isocausality introduced in \cite{PII}. We will 
encounter several mathematical problems to be faced, such as the 
generic infinite-dimensionality of the vector space we define, and the 
absence of a truly Lie algebra, or Lie group, structure. These 
impediments are briefly explained in this introduction.

Symmetries on a spacetime are often described by a group $G$ acting
on the underlying manifold and such that some adequate property---the 
metric, the null cone, the connection, etcetera--- is 
preserved by this action. Therefore every action gives rise 
to a group of transformations of the manifold $V$ which is a subgroup of the 
diffeomorphism group. Usually, all the essential information of the 
symmetry group is contained in the so-called {\em infinitesimal generators} 
of the group (defined in a way we briefly review in section 
\ref{Continuous}) which are vector fields on the manifold obeying an 
appropriate differential condition ---generally constraining the
Lie derivative of certain geometric objects along these vector fields---. 
Obvious examples are the Killing and conformal Killing vectors for isometries 
or conformal motions, respectively. Quite frequently, this set of vector fields 
form a Lie algebra under the Lie bracket operation $[\ ,\ ]$. Its 
dimension, however, can be finite or not:

{\em Finite dimensional Lie algebra.}
In this case the Lie algebra is isomorphic to the intrinsic Lie algebra of 
a finite-dimensional Lie group $G$, the transformation group being a 
realization of $G$ over the manifold. All this is well known 
and can be found in many textbooks (see e. g. \cite{MORITA,YANO,CHERN,KRAMERS}).
Actually, finite dimensional 
Lie groups have been studied for more than one hundred years 
and their main features as well as the general theory are already widely known.
Thus, it is possible to use all the machinery of Lie group theory as it has been 
largely done for example with isometries in order to find, classify and study 
solutions of four-dimensional Einstein's field equations 
\cite{KRAMERS}. Similar analyses 
have been performed for conformal motions, affine collineations, linear 
collineations and conformal collineations (see 
\cite{YANO,BARNES,FRWHALL,HALLDACOSTA,HALLLONIE,COLEY,HALLSTEELE,KRAMERS} and 
references therein) with perhaps a less firmly established physical 
interpretation. An account of the classical 
symmetries studied in General Relativity can be found in \cite{KATZIN,KRAMERS}.

{\em Infinite dimensional Lie algebra.}
Some examples of such Lie algebras are provided, in particular 
situations, by the generators of Riemann tensor collineations \cite{RIEMANN}
or of 2-dimensional conformal motions. The infinite-dimensional case
is still an open subject since there is no known general 
scheme to deal with it, hence, it remains 
highly unexplored apart from isolated attempts 
\cite{TSAMPARLIS,ZAFIRIS,sergi,KERR-SCHILD}. 
Nevertheless, we quote a first 
classification of infinite-dimensional Lie algebras 
presented in \cite{KAC}:      
\begin{enumerate}
\item Case of vector fields on 
differentiable manifolds. This is the typical case in General 
Relativity.  The main result is the classification of all 
simple Lie algebras performed by Cartan \cite{CARTAN}.     
\item Smooth mappings from differentiable manifolds onto finite 
dimensional Lie groups. Since Lie groups can be realized as groups of 
matrices, here we deal with matrices whose elements are smooth functions. 
An outstanding example is the group of rigid motions of $\r^n$ 
(\cite{REF} and references therein)
where the manifold is $\r$ and the Lie group is the semidirect 
product of $SO(n)$ with $\r^n$.
\item Linear operators in Hilbert spaces,
the common case in Quantum Mechanics.
\item Kac-Mody algebras.
\end{enumerate}

We will be concerned with (i), groups acting on 
differentiable manifolds. To get a flavour of the 
differences between finite and infinite-dimensional Lie algebras, recall 
that for a finite-dimensional Lie group $G$, 
the orbits of the action are differentiable submanifolds of the 
manifold, and sometimes one can construct these orbits explicitly. 
Furthermore, the generating vector fields are smooth complete vector 
fields. This result has a converse (Palais'
theorem, see \cite{PALAIS}). Contrarily, if the group $G$ is infinite 
dimensional, the previous nice properties no longer hold,
see \cite{HALLSYM} for a summary.             

This work is an extension and a development of the ideas put forward 
in \cite{LETTER} where we defined a symmetry transformation (called causal 
symmetry) which generalizes the well known case of conformal symmetries. 
The basic requisite for a transformation to be a causal symmetry is that
all future-directed vectors must be push-forwarded to future-directed 
vectors, so causal symmetries are simply causal mappings (see \cite{PII} 
for an explanation of this terminology) in which both the domain and 
the target manifolds are the same. A similar transformation
was used in \cite{LOW,H2} with other aims. These authors use the name 
``causally decreasing diffeomorphisms'' for our causal symmetries
and they study some of their properties under certain 
restrictions. We address the subject in its full generality ---recovering 
results in \cite{LOW,H2}--- and this is the reason why we have
preferred to keep our terminology and notation. 
Nonetheless we will indicate which of our results were already 
presented in those references.     

The main difference between causal symmetries and the above outlined 
more typical transformations is that the former do not form a group. 
Rather, they constitute a {\it monoid} of transformations 
which means that the inverse of a causal symmetry is not in general a causal 
symmetry. As a matter of fact, the only causal symmetries whose 
inverse is also a causal symmetry are the conformal transformations, 
which are thus included and perfectly identifiable within the general set of 
causal symmetries. Nevertheless, the set of causal symmetries is a 
{\em submonoid} of the diffeomorphism group of the manifold, or of one of 
its subgroups, so we can still apply some of the techniques and 
results used when dealing with groups of transformations.

Submonoids and their generalizations, called semigroups, are far from new 
structures and constitute a branch of research on their own (the difference 
between submonoids and semigroups is that semigroups have no identity 
element). A thorough study of these structures can be found in 
\cite{SEMIGROUP,HILGERT,LJAPIN}. They play, for instance, a key role  
 in the study of one-parameter semigroups of linear operators in Banach spaces
(see e.g. \cite{LINEAR-SEMIGROUP}). There are also examples of the 
relevance of submonoids of transformations in physics. For instance, in 
Quantum Mechanics the general time evolution of a state or a density matrix 
is ruled, under certain assumptions, by the so-called dynamical maps. 
The inverse of a dynamical map is not a dynamical map ---unless the map 
transforms pure states in pure states---. This is quite logical since there 
can be no evolution from a mixed state, which is always the outcome of a 
measure, to a pure state, which are states existing before any 
measure takes place, see \cite{DINAMICAL-MAP}. In an analogous manner, 
the causal symmetries we define, and their infinitesimal counterpart, 
will select or prefer a direction of time, the ``future'', making it 
thus impossible that the inverse transformations---necessarily 
leading to the ``past''--- be causal symmetries. This seems 
reasonable from a physical point of view, but involves the previously 
mentioned mathematical difficulties: submonoid structures and 
infinite dimensionality of the generating vector spaces.

Most of the material presented in this paper relies in the 
research carried out in \cite{PI} and \cite{PII} (henceforth PI and PII
respectively). A summary of them together with some new
results dealing with causal tensors have been placed in appendix A
which should be first studied by the reader not acquainted with these 
matters. We will refer either to this appendix or to any of those references when some 
of the results stated there is required. Part of our notation is also defined in 
this appendix although most of it was already used in PI or PII. 

The plan of the paper is as follows: in section \ref{causal-symmetries} 
we introduce causal symmetries, their principal null directions, and some of 
their basic properties. Section \ref{Continuous} is devoted to the study of 
one-parameter submonoids of causal symmetries, their classification in terms 
of canonical null directions and the definition of invariant subsets of the 
manifold with the same number of null directions. The necessary and sufficient 
conditions which the generating vector fields of causal symmetries must meet 
are found in section \ref{teoria-infinitesimal} giving the most general 
equation involving the Lie derivative of the metric tensor with respect to 
these generators. Finally, constants of motions and currents are constructed 
for causal symmetries in section \ref{conservedquantities}. 
Some examples are provided throughout. We close the paper with some 
conclusions.

\section{Causal symmetries}
\label{causal-symmetries}
Let us first of all set the notation used in this work. $V$ will stand for a differentiable Lorentzian  
$n$-dimensional manifold with appropriate smooth structures and metric $\rmg_{ab}$ 
(the signature convention is $(+,-,\dots,-)$). The further condition of 
analyticity will be required for some results.  Round and square 
brackets enclosing indexes denote symmetrization and antisymmetrization 
respectively. The tangent space at a point $x\in V$, $T_x(V)$, gives rise
to the tangent bundle $T(V)$ and cotangent bundle $T^{*}(V)$ in the standard
fashion as well as the bundle $T^{r}_s(V)$ of $r$-contravariant
 and $s$-covariant tensors formed from these two by means of the tensor product
$\otimes$. Boldface arrowed (un-arrowed) characters will represent 
contravariant (covariant) objects reserving the small letters for the specific
case of vectors and 1-forms. We make no distinction neither in our notation 
nor in our terminology between the elements of the previously cited bundles and their sections unless otherwise 
stated. In these cases, the value of the section $\u$ at the point 
$x$ shall be denoted by $\u|_x$. Vectors and 1-forms are related by the usual rule 
${\bf w}=\G(\cdot,\vec{\bf w})$ written in index notation as 
$w_a=\rmg_{ab}w^{b}$ so we can translate all definitions given for
contravariant objects into a covariant form and vice-versa (we will always use 
the same kernel letter for a vector and its associate 1-form).
By means of a suitable raising or lowering of indices, every rank-2 tensor can be 
brought into an element of $T^1_1(V)$ which can be regarded 
as an endomorphism on $T(V)$. We can do this as $T^a_{\ b}$ and $T_{a}^{\ b}$
 for $T_{ab}$ being both endomorphisms equivalent in
the symmetric case $T_{ab}=T_{ba}$. 
Therefore when we speak about the eigenvectors and eigenvalues of any rank-2 tensor
we will mean those of the mixed 1-1 tensor $T^a_{\ b}$.

For a diffeomorphism $\f:V\rightarrow W$ the push-forward
and pull-back are written as $\f'$ and $\f^{*}$ respectively. For later use 
we recall the following formula involving these operations: if $u^a(x)$ and 
$v_a(x)$ are differentiable sections of $T(V)$ and $T^{*}(V)$ respectively 
expressed in a local coordinate chart, and $y=\f(x)$ a diffeomorphism of $V$
in these coordinates, then the pull-back of the scalar $k(x)=v_a(x)u^a(x)$ is
\be
(\f^{*}k)(x)=k(\f(x))=(\f^{*}v)_b(x)(\f'^{-1}u)^b(x)
\label{ss-cc}
\ee
with the obvious generalization for higher rank tensors. The smooth
diffeomorphisms acting on $V$ form an infinite dimensional
continuous group denoted by $\Diff(V)$. 

We present next the main objects in this work. These 
are causal relations in which both the
domain manifold and the target manifold are the same Lorentzian
manifold $(V,\G)$ (see appendix A for notation and explanations).
\begin{defi}
For a Lorentzian manifold $(V,\G)$ we define $\C(V,\G)$ as the set of
 diffeomorphisms $\f$ such that $V\prec_{\f} V$. The elements of
 $\C(V,\G)$ are called {\em causal symmetries}.
\label{CV}
\end{defi}

\medskip
\noindent
{\bf Remark.} As shown in appendix A a diffeomorphism $\f\in\C(V,\G)$ iff 
$\f^*\G\in\DP^+_2(V)$ (theorem \ref{basic-prop2}) or, 
equivalently, iff $\f^*\G(\k,\k)\geq 0$ for all 
$\k\in\partial\Theta_{x}$, $\forall x\in V$ (theorem \ref{nullconvergence}). 

According to the comments of appendix A, the set $\C(V,\G)$ contains as a
proper subset the group of conformal symmetries of our manifold which
we denote by $\Conf(V,\G)$. The
general properties of causal relations as well as the relationship
between causal and conformal relations discussed in appendix A
 translates also to the set of causal symmetries as follows 
(see proposition \ref{MUFI} for the definition of $\mu(\f)|_x$):
\begin{prop}
The next properties hold
\begin{enumerate}
\item $\f_1,\f_2\in\C(V,\G)$ $\Rightarrow$ $\f_1\circ\f_2\in\C(V,\G)$.
\item $\Conf(V,\G)=\C(V,\G)\cap\C(V,\G)^{-1}$.
\item $\f_1,\f_2\in\C(V,\G)\Rightarrow\mu(\f_1\circ\f_2)|_x\subseteq\mu(\f_2)|_x$. 
\end{enumerate}
\label{null property}
\end{prop}
\P Properties {\it i)} and {\it ii)} are straightforward consequences
of the analog properties for causal relations 
(proposition 3.3 of PII) and so we will not
repeat the proof here. Property {\it (iii)} follows from the equation
$$
0=(\f_2^{*}\f_1^{*}\G)|_x(\k,\k)=(\f_1^{*}\G)|_{\f_2(x)}(\f'_2\k,\f'_2\k)\
\ \forall\k\in\mu(\f_1\circ\f_2)|_{x},
$$
because the null vector $\k$ remains null under $\f'_2$ 
(see proposition \ref{MUFI}).\N

This proposition tells us that $\C(V,\G)$ with the inner operation of
diffeomorphism composition is a submonoid of $\Diff(V)$. Recall that a
set $S$ with an inner operation is a monoid if the operation is
associative and there exists an
identity element for this operation in $S$ 
(if such identity does not exist $S$ is
called a semigroup). Furthermore, if $S\subset G$ is a {\it proper
submonoid} of $G$ we define its group of units $H(S)$ as $S\cap
S^{-1}$. The previous proposition tells us what is the group of units of 
$\C(V,\G)$. 
\begin{prop}
The group of units of $\C(V,\G)$ is $\Conf(V,\G)$.\N  
\label{UNITGROUP}
\end{prop}
An interesting property of the group of units is
that it is the maximal subgroup contained in $S$ in the sense that
there is no other bigger subgroup of $G$ contained in $S$ possessing
$H(S)$ as a proper subgroup. See \cite{SEMIGROUP} for the proof of
this and other properties of submonoids and semigroups. The causal symmetries
not being conformal transformations deserve thus a special name.
\begin{defi}
Any $\f\in \C(V,\G)\!\!\setminus\!\!\Conf(V,\G)$ is called a {\em proper}
causal symmetry.  
\label{PROPER}
\end{defi}
As is clear $\C(V,\G)$ depends on the background metric $\G$ chosen for our
Lorentzian manifold. Nonetheless some properties of $\C(V,\G)$ are shared by
different elections of the metric $\G$, in particular $\C(V,\G)$ is conformally 
invariant: 

\begin{prop}
$\C(V,\G)=\C(V,\sigma\G)$ for any positive smooth function $\sigma$ on $V$.
\label{CONFINVARIANT}
\end{prop}
\P This happens because a future tensor with respect to a
fixed background metric $\G$ will remain future with respect to every conformal metric $\sigma\G$ if
 $\sigma$ is positive, and this applies to $\f^*\G$, $\forall\f\in\C(V,\G)$.\N  

Another result connecting the causal structure of $(V,\G)$ as it was
defined in PII and the set of causal symmetries is the next (see 
definition \ref{isocaus-def} for the concept of {\em isocausality})
\begin{prop}
For two isocausal Lorentzian manifolds $(V,\G)$ and $(W,\tT)$, there is
 a one-to-one correspondence between the sets $\C(V,\G)$ and
$\C(W,\tT)$.
\label{correspondence}
\end{prop}
{\bf Remark.} Note that this is a nontrivial result as the
cardinality of the diffeomorphism group of each manifold need not be
the same as the cardinality of $\C(V,\G)$ or $\C(W,\tT)$.

\noindent
\P Under the hypotheses of this proposition we know that
$V\prec_{\psi_1}W$ and $W\prec_{\psi_2}V$. Now, it is rather simple to
check that for every $\f\in\C(V,\G)$,
$\psi_1\circ\f\circ\psi_2\in\C(W,\tT)$. Similarly every
$\chi\in\C(W,\tT)$ gives rise to an element of $\C(V,\G)$ defined as
$\psi_2\circ\chi\circ\psi_1$. These correspondences are injective so
that Bernstein equivalence theorem \cite{BERN} leads to the result. \N

If we consider now elements of $\DP^+_r(V)$ 
(see appendix A, definition \ref{CAUS-TENSR})
we know according to proposition
\ref{basic-prop} that they will remain in $\DP^+_r(V)$ under the pull-back of
every diffeomorphism of $\C(V,\G)$. The next proposition tells us that 
their principal directions also behave in a precise fashion when push-forwarded by the elements 
of $\C(V,\G)$ (see definition \ref{SIGMA} for the set $\s(\T)$).
\begin{prop}
If $\f\in\C(V,\G)$ then we have the following inclusions $\forall x\in V$:
\begin{enumerate}
\item $\f'[\mu(\f^{*}\T|_x)]\subseteq\mu(\T|_{\f(x)})$ $\forall\T\in\DP^+_r(V)$.
\item $\mu(\f^{*}\T|_x)\subseteq\mu(\f)|_{x}$ $\forall\T\in\DP^+_r(V)$.
\item $\forall$ $\T\in\DP^+_2(V)$, either 
\begin{enumerate}
\item $\T {}_{1}\times_{1}\T=0$ ($\Longrightarrow$
$\s(\f^*\T|_x)=\s(\T|_x)=\partial\Theta^+_{x}$),   
\item or $\s(\f^*\T|_x)\subseteq\mu(\f)|_x$ and 
$\f'[\s(\f^{*}\T|_x)]\subseteq\s(\T|_{\f(x)})$.
\end{enumerate} 
\end{enumerate}
\label{fin-trans}
\end{prop}
\P The first two pints follow from 
$$
0=\f^*\T|_x(\k,\dots,\k)=\T|_{\f(x)}(\f'\k,\dots,\f'\k),
$$
where $\k\in\mu(\f^*\T|_x)$ (note that $\f'\k|_{\f(x)}$ 
is then null for every $\k|_x\in\mu(\T|_x)$). To prove (iii),
from the comments after definition \ref{SIGMA} one knows that 
every $\k_1\in\s(\f^*\T|_x)$ is either in $\mu(\f^*\T|_x)$ or, 
otherwise, there is another null $\k_2$ (given by $k_{2}^b=T^b{}_{c}k^c_{1}$)
such that 
\be
0=\f^{*}\T|_x(\k_2,\k_1)=\T|_{\f(x)}(\f'\k_2,\f'\k_1). \label{mig}
\ee
In the former case the statement is a subcase of point (i). 
Suppose then that $\k_1\notin\mu(\f^*\T|_x)$. From (\ref{mig})
we deduce, $\T$ being a future tensor, that
at least one of $\f'\k_1|_{\f(x)}$ or $\f'\k_2|_{\f(x)}$ must be 
null. If both of them are null, (b) is proven. If $\f'\k_1|_{\f(x)}$ were null 
but $\f'\k_2|_{\f(x)}$ were timelike, (\ref{mig}) would imply 
that in fact $\T|_{\f(x)}(\cdot \, ,\f'\k_1)=0$ so that 
$\f^{*}\T|_x(\cdot \, ,\k_1)=0$ and thus $\k_{1}$ would be in 
$\mu(\f^*\T|_x)$ against the assumption. 
Finally, if $\f'\k_2|_{\f(x)}$ is null but $\f'\k_1|_{\f(x)}$ is 
timelike then, as before, $\f^{*}\T|_x(\k_2,\cdot \, )=0$ whence
$\k_{2}\in \mu(\f^*\T|_x)\subseteq\mu(\f)|_{x}$ and, furthermore,
$\f^{*}\T|_x = \K_{2}\otimes \mathbf{v}$ for some 
$\mathbf{v}\in\DP^{+}_{1}|_x$. It follows that $\T=\K_{2}\otimes 
\mathbf{u}$ for a future-directed causal $\mathbf{u}$, in which case, 
as is clear, $\T {}_{1}\times_{1}\T=0$ and every null vector belongs to 
$\s(\f^*\T|_x)=\s(\T|_x)$. \N

A first consequence of this result is that $\mu(\f^{*}\T|_x)$ will be 
empty if either $\mu(\f)|_x$ itself is empty or
$\mu(\T|_{\f(x)})=\emptyset$. Similar considerations hold for 
$\s(\f^*\T|_x)$. Unless for the particular case of tensors 
$\T {}_{1}\times_{1}\T=0$, part of the above results can be gathered in the chain 
$\mu(\f^*\T|_x)\subseteq\s(\f^*\T|_x)\subseteq\mu(\f)|_x$, which is a property 
to be used later on, and that can be refined if we impose further algebraic 
constraints upon $\T$ when it has rank 2.
\begin{prop}
If the symmetric $\T\in\DP^+_2(V)$ has Lorentzian signature at every 
point of the manifold 
then $\s(\f^*\T|_x)=\mu(\f^*\T|_x)$, $\forall x\in V$ and every $\f\in\C(V,\G)$. 
\label{signature}
\end{prop}
\P The signature of $\f^*\T|_x$ is the same as the signature of 
$\T|_x$ so the result follows from application of proposition \ref{lorentzian} to the tensor 
$\f^*\T|_x$.\N

An immediate consequence of this is
$$\s(\f^*\G)=\mu(\f^*\G)=\mu(\f), \,\,\,\,\, \forall \f\in\C(V,\G).$$  

The tensor $\f^{*}\G|_x$ may change its algebraic character from point to point
so the number of independent canonical null directions will in 
general be different depending on the point of $V$. Therefore, it is quite
reasonable to collect in sets the points of the manifold with the 
same number of these directions. 
\begin{defi}
Let $\f$ be an element of $\C(V,\G)$ and consider the future tensor
$\f^{*}\G$. Then we define the following 
sets:
\begin{eqnarray*}
\hspace{-2cm} \Nm^m_{\f}=\{x\in V: \f^{*}\G|_x \ \mbox{has $m$ linearly 
independent null eigenvectors}\},\\ 
\hspace{-1cm} \Um_\f=\{x\in V:\mu(\f^{*}\G|_x)=\emptyset\}.
\end{eqnarray*}
The union of the sets $\Nm^m_\f$ for all the values of $m$ is denoted by
$\Nm_\f$.
\label{decom-segre}
\end{defi}
The set $\Nm^m_{\f}$ is formed by the points $x\in V$ at
which $(\f^{*}\G)|_x$ has Segre type
$[\overbrace{(1,\ 1\dots 1)}^{m}1\dots 1]$ and its spatial degeneracies 
(or $[2\ 1\dots 1]$ and its degeneracies if $m=1$). The allowed Segre
types for a rank-2 causal tensor are summarized in table \ref{Segre}
of appendix A.

Following \cite{RAUL}, whenever we work with a smooth section of $\DP^+_2(V)$, 
we can decompose our spacetime $V$ disjointly in sets defined by 
having a constant Segre type of the section values at their interior,
plus a remainder $X$ with no interior which is closed in $V$. 
This is an application of theorem 1 of \cite{RAUL} which states that a section
of the bundle $T^0_2(V)$ of symmetric rank-2 tensors on a four-dimensional 
Lorentzian manifold gives rise to a decomposition of $V$ in the following
disjoint subsets:
\bnr
\fl V=X\cup [1,111]\cup[1,1(11)]^{\circ}\cup[(1,1)11]^{\circ}
\cup[(1,1)(11)]^{\circ}
\cup[(1,11)1]^{\circ}\cup[1,(111)]^{\circ}\cup\\
\fl[(1,111)]^{\circ}\cup[211]^{\circ}\cup
[2(11)]^{\circ}\cup[(21)1]^{\circ}\cup[(211)]^{\circ}\cup
[31]^{\circ}\cup[(31)]^{\circ}\cup[z\bar{z}11]\cup[z\bar{z}(11)]^{\circ},
\enr
where the Segre notation is also used to mean each of the sets whose interior
has constant algebraic type. If the section is a causal tensor then we must only 
take into account its allowed Segre types, shown in table \ref{Segre}. 
For the section $\f^*\G$ we thus have that
these sets are related to $\Um_\f$ and $\Nm^m_{\f}$ in the following way for 
dimension four:
\bnr
\fl \Um_{\f}=[1,111]\cup[1,1(11)]\cup[1,(111)],\ \ 
\Nm^1_{\f}=[211]\cup[2(11)]\cup[(21)1]\cup[(211)],\\
\Nm^2_{\f}=[(1,1)11]\cup[(1,1)(11)],\ \ 
\Nm^3_{\f}=[(1,11)1],\ \ \Nm^4_{\f}=[(1,111)].
\enr 
Thus the splitting of $V$ in terms of these sets looks like:
\be
 V=\Um_{\f}^{\circ}\cup(\Nm^1_{\f})^{\circ}\cup(\Nm^2_{\f})^{\circ}
\cup(\Nm^3_{\f})^{\circ}\cup(\Nm^4_{\f})^{\circ}\cup X.\ \ 
\label{SPLITTING}
\ee
Of course this splitting depends on the element $\f$ of $\C(V,\G)$ under
consideration and we may have different splittings of our manifold for
the different elements of $\C(V,\G)$. This is of no relevance here but it
will play a role when we study continuous causal symmetries in the next
section where we will show that the resulting sets defined by means of
these splitting techniques have nice invariant properties. We must also add
that, as the authors of \cite{RAUL} argue at the end of their paper, their result
can also be extended to dimensions greater than four so we will assume henceforth 
the obvious
$n$-dimensional generalization of (\ref{SPLITTING}). 

\section{Continuous theory}
\label{Continuous}
As briefly explained in the Introduction, 
when one studies actions of symmetry groups on a spacetime 
such as isometries or conformal motions, one uses the intimate 
relationship between the group of transformations itself and its Lie 
algebra of ``infinitesimal generators''. This can be done in fact
not only for the finite dimensional cases, but also for the Lie 
algebras of infinite dimension. In any case, one goes from the global 
theory to the local one by introducing local one-parameter groups of
diffeomorphisms belonging to the transformation group under 
consideration. These local one-parameter groups of transformations are denoted
by $\{\f_s\}_{s\in I}$ where $s$ is the canonical parameter, $I$ is an interval 
of the real line containing $0$ and $\f_0=Id$. The possibility of working with 
global one-parameter groups with $I=\r$ or $S^1$ depends on the group under 
consideration although this is irrelevant to our purposes. 
A vector field $\xiv$ is said to
be an infinitesimal generator of the symmetry group (a {\em generating
vector} in short) if there exists one of the above defined
one-parameter groups $\{\f_s\}_{s\in I}$ such that in
local coordinates $\{x^a\}$:
\[
\xi^a(x)=\left.\fr{d\f^a_s(x)}{ds}\right|_{s=0}\ \forall x\in V.
\]
The set of generators form a Lie algebra under the usual Lie bracket of
vector fields. Sometimes it is possible to find a differential
condition which determines all the generating
vectors of a certain type of symmetry as for instance $\lie\G=0$ or
$\lie\G=\alpha\G$ for isometries and conformal motions,
respectively. Therefore, the knowledge of such differential conditions
may provide full information about the symmetry under study, so 
its search for the causal symmetries is one of our main tasks from now 
on.

Nevertheless, an immediate difficulty arises: when dealing with the 
causal symmetries we are confronted with the problem that $\C(V,\G)$ is not 
actually a group. However, $\C(V,\G)$ keeps group-like structures, 
hence we may expect that some of the above mentioned properties of other
symmetries will also be maintained for the causal symmetries. To start 
with, we already know that $\C(V,\G)$ has the algebraic structure of a 
monoid and this allows us to say a word about any topological group with 
a subset liable to be realized by means of elements of $\C(V,\G)$. 
Of course, we will be mainly concerned with the nontrivial case where
those elements are proper causal symmetries so that $\C(V,\G)\setminus 
\Conf(V,\G)$ should be nonempty. In this case, if $G$ is 
the mentioned topological group and $S\subset G$ is the set which gives rise to 
$\C(V,\G)$ then $S$ cannot be a compact subset of $G$ 
(see \cite{SEMIGROUP}, pag 370).
Moreover, and despite $\C(V,\G)$ not being a group, we can still define 
one-parameter groups $\{\f_s\}_{s\in I}$ such that they have elements in 
$\C(V,\G)$. If we call $J\subset I$ the subset of $I$ such that
$\{\f_s\}_{s\in J}\subset\C(V,\G)$ then we get the following
relationship between $J$ and $I$ if $J\supseteq[0,\epsilon)$ for
$\epsilon>0$.

\begin{prop}
If $\{\f_s\}_{s\in I}\cap\C(V,\G)\supset\{\f_s\}_{s\in\bar{I}}$ with
$\bar{I}=[0,\epsilon)$ then $\{\f_s\}_{s\in\r^+\cap I}\subset\C(V,\G)$.
\label{generatriz}
\end{prop}
\P This is a straightforward consequence of the fact that every real
number in $I$ can be written as a finite sum of numbers in
$\bar{I}$.\N

Therefore, we deduce that under the above conditions
$\{\f_{s}\}_{s\in\r^+\cap I}$ is itself a {\it local submonoid of
causal symmetries} (we will sometimes drop the tag ``local'' for 
the sake of brevity). We will work with these 
structures upon which we will try to carry out the generalization
program mentioned before.

\begin{prop}
Let $\{\f_s\}_{s\in\r^+\cap I}$ be a local submonoid of causal symmetries
and suppose that the diffeomorphism $\f_{s_0}$ for $|s_0|\in\r^+\cap
I$ is a conformal transformation. Then $\{\f_{s}\}_{s\in I}$ is a
local one-parameter group of conformal transformations.
\label{group}
\end{prop}
\P For any null vector $\k$ we have according to theorem \ref{CONF}
that $\f'_{|s_0|}\k$ must be also null since $\f_{|s_0|}$ is
conformal. If we write this conformal transformation as 
$\f_{|s_0|}=\f_{s_1}\circ\f_{s_2}$, $s_1, s_2>0$, we immediately get, 
using proposition \ref{caus-trans}, that $\f'_{s_2}\k$ must be null 
from what we deduce that $\f'_s\k$ must be null if $0\leq s\leq s_0$. 
Since this has been done for an arbitrary null vector, we 
conclude using again theorem \ref{CONF} that $\f_{s}$ are conformal 
transformations for $0\leq s\leq s_0$ and hence $\forall s\in I$ 
in view of the group properties of conformal transformations.\N 
     
This proposition tells us that one-parameter submonoids of causal symmetries
either consist exclusively of conformal transformations (and they can be 
extended to one-parameter groups recovering the classical theory) or they 
contain just the identity as the unique conformal transformation. 
We will henceforth study one-parameter submonoids with the identity as the only 
conformal transformation.

\begin{coro}
There are no realizations of $S^1$ as a one-parameter submonoid of 
proper causal symmetries. 
\end{coro}
\P We can prove this by 
noting that if $S^1$ were realized as a one-parameter submonoid of proper 
causal symmetries then the parameter $s$ labelling each diffeomorphism
 could be chosen running in the interval $[0,2\pi]$ with $Id=\f_0=\f_{2\pi}$.
 Proposition \ref{group} would imply then that $\f_s$ 
would be a conformal transformation $\forall s\in[0,2\pi]$.\N

We can put this result another way by saying that proper causal symmetries 
cannot have cyclic orbits.

\subsection{Canonical null directions of submonoids of causal symmetries}
We will extend now the study of the set of canonical null directions in 
the previous section to the case of local
 continuous one-parameter submonoids of causal
symmetries $\{\f_s\}_{s\in I\cap\r^+}$. To start with, we prove 
one of our fundamental results. 
\begin{theo}
If $\{\f_s\}_{s\in I\cap\r^+}\subset\C(V,\G)$ and $\G$ is analytic we have 
the property
$\mu(\f^{*}_{s_0}\G|_x)=\mu(\f^{*}_{s_1}\G|_x)$ $\forall s_0,\ s_1>0$.      
\label{corner-stone}
\end{theo}
\P If $s_0>0$, proposition \ref{null property} implies that 
$\forall s_1\in[0,s_0]$,
 $\mu(\f_{s_0})|_x=\mu(\f_{s_0-s_1}\circ\f_{s_1})|_x\subseteq\mu(\f_{s_1})|_x$  
and hence $\mu(\f^{*}_{s_0}\G|_x)\subseteq\mu(\f^{*}_{s_1}\G|_x)$.
Now if $\k\in\mu(\f_{s_1}^{*}\G|_x)$ and we define the
function $f_{\k}(s)\equiv\f^{*}_s\G|_x(\k,\k)$ with $\k$ fixed, we deduce that 
$f_{\k}(s)=0$ for every $s\in I$ since this is an analytic function vanishing 
on the interval $[0,s_1]$ (due to the previous stated inclusion).
This means that 
$\k\in\mu(\f^{*}_{s_0}\G|_x)$ from what the inclusion 
$\mu(\f_{s_1}^{*}\G|_x)\subseteq\mu(\f^{*}_{s_0}\G|_x)$ follows.\N

\smallskip
\noindent
{\bf Remarks.}
\begin{enumerate}
\item The proof of this proposition implies in fact that, for an analytic metric
$\G$, $0=\f^*_s\G|_x(\k,\k)$ $\forall s\in I$ if $\k\in\mu(\f^*_{s_0}\G|_x)$ 
for a certain $s_0>0$. 

\item Observe that the elements of $\mu(\f^*_s\G)$, $s>0$ are the null vectors
$\k$ which remain null under the action of $\{\f_s\}_{s\in I}$, that is, such that
$\f'_s\k$ is null for all $s\in I$. Thus, this can be taken as 
definition of {\em canonical null directions} of the submonoid.
\end{enumerate}       

We have just proven that the principal null directions of $\f^{*}_s\G$ do not 
depend on the value of $s>0$ provided $\G$ is analytic. Nonetheless the algebraic
type of $\f^{*}_s\G$, and the number of such directions, might change from 
point to point and so we can define the sets $\Nm_{\f_s}^m$, $\Um_{\f_s}$ and 
$\Nm_{\f_s}$ which split the manifold in sets with a constant number 
of them for the tensor $\f^{*}_s\G$. The previous proposition can be rewritten 
in terms of these sets as 
$\Nm^m_{\f_{s_1}}=\Nm^m_{\f_{s_2}}\equiv\Nm^m,\ {\cal N}_{\f_{s_1}}=
{\cal N}_{\f_{s_2}}\equiv{\cal N},\ 
\Um_{\f_{s_1}}=\Um_{\f_{s_2}}
\equiv\Um$, $\forall s_1, s_2\in I\cap\r^+$ but we can even say more about 
these sets.
\begin{prop}
If $\G$ is analytic then for every one-parameter submonoid of causal 
symmetries $\{\f_s\}_{s\in I\cap\r^+}$ we have that
$$
\f_{s}(\Um)=\Um,\ \f_{s}(\Nm^m)=\Nm^m,\ \f_s({\cal N})={\cal N}\ 
\forall s\in I.
$$
\label{invariance}
\end{prop}
\vspace{-.5cm}
\P To prove this proposition we must show that the set of 
canonical null directions is preserved under the flow of $\{\f_s\}_{s\in I}$, 
that is to say, they are the same for the tensors $(\f^{*}_{s_0}\G)|_x$ 
and $(\f^{*}_{s_0}\G)|_{\f_s(x)}$ 
(notice that in view of theorem \ref{corner-stone} the precise $s_0$ has 
no relevance here as long as it is kept positive). Thus, we are going
to establish a bijection between the set of null eigenvectors
of $(\f^{*}_{s_0}\G)|_x$ and those of $(\f^{*}_{s_0}\G)|_{\f_s(x)}$. 
Let us start by writing down the equation
\be
\fl(\f^{*}_{s_0}\G)|_{\f_{s}(x)}(\f'_{s}\u_1,\f'_{s}\u_2)=
(\f^{*}_{s_0+s}\G)|_{x}(\u_1,\u_2),\ \ \forall \u_1,\u_2\in T(V),\ s\in I,
\label{bb}
\ee  
and choosing any null vector $\k\in\mu(\f^{*}_{s_0}\G|_x)$. Theorem
\ref{corner-stone} implies that $\k\in\mu(\f^{*}_{s_0+s}\G|_x)$ for $s>0$ and 
hence use of (\ref{bb}) tells us that $\f'_{s}\k\in\mu(\f^*_{s_0}\G|_{\f_s(x)})$
which means that $\f'_s$ sets an injection between $\mu(\f^{*}_{s_0}\G|_x)$ and
$\mu(\f^*_{s_0}\G|_{\f_s(x)})$. To show that $\f'_s$ is also onto, we pick any
$\u\in\mu(\f^{*}_{s_0}\G|_{\f_{s}(x)})$ and  
look again at the proof of theorem \ref{corner-stone} deducing that the 
vector $\k$ such that $\f'_{\bar{s}}\k=\u$ must also be null 
$\forall\bar{s}\in I\cap\r^+$ so that inserting
this expression of $\u$ into equation (\ref{bb}) we conclude that 
$0=(\f^{*}_{s_0+\bar{s}}\G)|_x(\k,\k)$ which means according to theorem 
\ref{corner-stone} that $\k\in\mu(\f^{*}_{s_0}\G|_x)$. Therefore the push-forward
$\f'_{s}$ sets the sought bijection between 
$\mu(\f^{*}_{s_0}\G|_{x})$ and $\mu(\f^{*}_{s_0}\G|_{\f_{\bar{s}}(x)})$ which 
leads us to the required invariance properties of $\Um$ and $\Nm^m$
for $s>0$ and by extension $\forall s\in I$ due to the group
property of $\{\f_s\}_{s\in I}$ (for example, 
$\f_{s}(\Um)=\Um\Rightarrow \Um=\f_{-s}(\Um)$.)\N  

\smallskip
\noindent
{\bf Remarks.}
\begin{enumerate}
\item $\forall x\notin X$ (see (\ref{SPLITTING})), its orbit  
${\cal O}_x=\{y\in V:\f_s(x)=y\}$ is a subset of one of the invariant sets 
$\Nm^m$ or $\Um$.
\item Theorem \ref{corner-stone} and proposition \ref{invariance} 
allow us to speak about the canonical
null directions of a one-parameter submonoid of causal symmetries independently
of the parameter $s$ or of the region of the manifold (once we have chosen one of 
the sets $\Um$ or $\Nm^m$). We will work in the sequel 
with manifolds with just a single component $\Um$ or $\Nm^m$ and the metric 
tensor will be assumed analytic there unless otherwise stated. 
By reasons which shall be clear later, we will denote the set of canonical null 
directions at a point as $\mu_{\xiv}|_x$ being $\xiv$ the generating vector 
field of $\{\f_s\}_{s\in I}$. As always we may give the global 
definition of this as $\mu_{\xiv}=\bigcup_{x\in V}\mu_{\xi}|_x$ 
which is a subset of the bundle $T(V)$ and we will use the same 
symbol if there is need of considering sections over $\mu_{\xiv}$. 
Theorem \ref{corner-stone} tells us that 
\be
\f'_s(\mu_{\xiv})=\mu_{\xiv},\  \forall s\in I.
\label{invte-mu}
\ee
 As happened with the set of principal directions, $\mu_{\xiv}|_x$ is not a 
vector space but we can pick up a maximum number of linearly independent 
elements denoted by $\chi(\mu_{\xiv}|_x)$ which is simply 
dim$Span(\mu_{\xiv}|_x)$.
\item  The maximum number of linearly independent canonical null directions can be used as a first
mean to classify these symmetries as they were used to classify causal relations
in PII. The case with $\chi(\mu_{\xiv}|_x)=n$, $\forall x\in V$ 
is given by the conformal transformations of the manifold $V$  
whereas the case with $\chi(\mu_{\xiv}|_x)=m<n$, $\forall x\in V$ shall be 
called partly conformal case and its elements {\it $\fr{m}{n}$-partly conformal 
symmetries} as they do not preserve the full null cone. 
The most simple case of nontrivial partly
conformal symmetries is the one with $m=n-1$ (we will give
in proposition \ref{n-1} a very simple differential condition for the generating vectors of 
these symmetries). The complexity increases as the number $m$ decreases. 
A one-parameter submonoid of causal symmetries with a single linearly independent 
canonical null direction is called {\it degenerate},
and {\it nondegenerate} otherwise.
\end{enumerate} 

The results of proposition \ref{fin-trans} can also be extended to the case of 
one-parameter submonoids of causal symmetries as follows:
\begin{prop}
If $\{\f_s\}_{s\in I\cap\r^+}\subset\C(V,\G)$, then 
$\f'_s[\mu(\f^{*}_s\T|_{x})]=\mu(\T|_{\f_s(x)})\cap\mu_{\xiv}|_{\f_s(x)}$ 
$\forall\T\in\DP^+_r(V)$.
\label{tr-T}
\end{prop}
\P The inclusion $\f'_s[\mu(\f^{*}_s\T|_{x})]
\subseteq\mu_{\xiv}|_{\f_s(x)}\cap\mu(\T|_{\f_s(x)})$ comes
from proposition \ref{fin-trans} (actually it is also true for
non-analytic metrics). If $\k$ is any null vector in 
$\mu_{\xiv}|_{\f_s(x)}\cap\mu(\T|_{\f_s(x)})$ we know that 
$\f'_{-s}\k\in\mu_{\xiv}|_{x}$ $\forall s\in I$
 and we have 
$0=\T|_{\f_s(x)}(\k,\dots,\k)=(\f^{*}_s\T)|_{x}(\f'_{-s}\k,\dots,\f'_{-s}\k)$ 
which means 
for $s>0$ that $\f'_{-s}\k\in\mu(\f^{*}_s\T|_{x})$ from what we conclude that 
$\f'_{-s}(\mu_{\xiv}|_{\f_s(x)}\cap\mu(\T|_{\f_s(x)}))
\subseteq\mu(\f^{*}_s\T|_{x})$
 leading to the other inclusion.\N

As an application, take any causal tensor 
$\T$ such that $\mu(\T|_x)\cap\mu_{\xiv}|_x=\mu_{\xiv}|_x$, $\forall x\in V$. 
This proposition implies then that $\mu(\f^{*}_s\T|_x)=\mu_{\xiv}|_x$, 
$\forall s\in I\cap\r^+$,
 $\forall x\in V$. The opposite case 
occurs when $\mu_{\xiv}=\emptyset$ which entails $\mu(\f^{*}_s\T|_x)=\emptyset$, 
$\forall x\in V$, $\forall s\in I\cap\r^+$ and for all $\T\in\DP^+_r$ 
(this case is true regardless of 
the analytic properties of the background metric). Another important outcome of 
this is the following.
\begin{prop}
If $\{\f_s\}_{s\in I\cap\r^+}\subset\C(V,\G)$ then for every $\T\in\DP^+_2(V)$
such that $\mu_{\xiv}|_x\subseteq\mu(\T|_x)$ and 
$\T{}_{1}\times_{1}\T\neq 0$ we have that 
$\s(\f^*_s\T|_x)=\mu(\f^*_s\T|_x)=\mu_{\xiv}|_x$, $\forall s>0$.
\label{S=M}
\end{prop}
\P This follows from the already proven inclusions 
$\mu(\f^*_s\T|_x)\subseteq\s(\f^*_s\T|_x)
\subseteq\mu(\f^*_s\G|_x)=\mu_{\xiv}|_x$, $s>0$ (see proposition 
\ref{fin-trans}) and the previous proposition.\N

\noindent
{\bf Remark.} Note that the property $\s(\f^*_s\T|_x)=\mu(\f^*_s\T|_x)$ 
holds with no assumptions on $\mu_{\xiv}|_x$ if $\T$ has 
Lorentzian signature (see proposition \ref{signature}).

The invariance property (\ref{invte-mu}) can be recast in a more handleable way. 
\begin{lem}
Let $\k^1,\dots,\k^m\in \mu_{\xiv}$ be linearly independent
with $m\equiv\chi(\mu_{\xiv})$. Then 
$\f'_s\k^j|_{\f_s(x)}\in Span\{\k^1|_{\f_s(x)},\dots\k^m|_{\f_s(x)}\}$ and
$\f^{*}_s\K^j|_{x}\in Span\{\K^1|_x,\dots\K^m|_x\}$
 $\forall j=1,\dots,m$, $s\in \r^+\cap I$.
\label{form1}
\end{lem} 
\P The invariance law (\ref{invte-mu}) says that 
$\f'_s(\mu_{\xiv}|_x)=\mu_{\xiv}|_{\f_s(x)}$ so for every
 $\k|_x\in\mu_{\xiv}|_x$ 
we have that $\f'_s\k|_{\f_s(x)}\in\mu_{\xiv}|_{\f_s(x)}$
which means that 
$\f'_s\k|_{\f_s(x)}\in Span\{\k^1|_{\f_s(x)},\dots,\k^m|_{\f_s(x)}\}$ since by 
assumption these vectors are a basis of the subspace generated by 
$\mu_{\xiv}|_{\f_s(x)}$. Now the property 
$\mu_{\xiv}|_x=\f'_{-s}(\mu_{\xiv}|_{\f_s(x)})$ leads us to a similar 
statement about $\f'_{-s}\k|_x$ if $\k|_{\f_s(x)}\in\mu_{\xiv}|_{\f_s(x)}$
and hence $\f'_{-s}\k|_x\in Span\{\k^1|_x,\dots,\k^m|_x\}$. This last result
is necessary 
to prove the property involving the differential forms because if
we pull-back the defining relation of each $\K|_x$ in terms of $\k|_x$ we get:
\be
k_a=\rmg_{ab}k^{b}\Rightarrow 
(\f^{*}_sk)_a=(\f^{*}_s\rmg)_{ab}(\f'_{-s}k)^b,
\ee
where an appropriate generalization of (\ref{ss-cc}) has been used. The sought 
result follows from the fact that  
$\f'_{-s}\k|_x$ ($s>0$) is a null eigenvector of $\f^*_s\G|_x$ 
which belongs to $Span\{\k^1|_x,\dots,\k^m|_x\}$.\N

It is clear that these and other forthcoming properties can be generalized 
to every $s\in I$ but for our present 
purposes it is enough to stick to positive values of $s$.
The relevance of these invariance properties shall be addressed later on. To 
that end we rewrite them in a more geometrical and compact way.
\begin{prop}
If $\{\k^1,\dots,\k^m\}$ and $\{\K^1,\dots,\K^m\}$, are the set of vectors 
and forms introduced in the previous lemma then the 
$m$-vector $\vec{\O}=\k^1\wedge\dots\wedge\k^m$ and the $m$-form 
$\O=\K^1\wedge\dots\wedge\K^m$ behave under the flow
of $\{\f_s\}$ as:
\be
\fl\f'_s\vec{\O}|_{\f_s(x)}=
\fr{\l^m_s(\f_s(x))}{\s_s(\f_s(x))}\vec{\O}|_{\f_s(x)},\hspace{1cm}
\f^{*}_s\O|_x=\s_s(x)\O|_x,\ \ s\in I\cap\r^+,
\label{omega}
\ee
where $\l_s(x)$ is the common eigenvalue of the elements of $\mu_{\xiv}$. 
\label{form2}
\end{prop}  
\P Under the hypotheses of lemma \ref{form1} we deduce by means of 
$\f'_s(\k^1\wedge\dots\wedge\k^m)|_{\f_s(x)}=
\f'_s\k^1|_{\f_s(x)}\wedge\dots\wedge\f'_s\k^m|_{\f_s(x)}$ and 
$\f^*_s(\K^1\wedge\dots\wedge\K^m)|_x=
\f^*_s\K^1|_x\wedge\dots\wedge\f^*_s\K^m|_x$
that $\f'_s\vec{\O}|_{\f_s(x)}\propto\vec{\O}|_{\f_s(x)}$ and 
$\f^*_s\O|_x\propto\O|_x$.
 We can freely set one of the proportionality factors appearing in
these  equations so we put 
$\f^*_s\O|_x=\s_s(x)\O|_x$. The other 
proportionality factor called $\beta_s$ is found by pulling back the equation 
$\Omega_{a_1\dots a_m}(x)=
\rmg_{a_1b_1}(x)\cdots\rmg_{a_mb_m}(x)\Omega^{b_1\dots b_m}(x)$ which leads us 
to
\be
\f^{*}_s\Omega_{a_1\dots a_m}(x)=
(\f^{*}_s\rmg)_{a_1b_1}(x)\cdots(\f^{*}_s\rmg)_{a_mb_m}(x)
(\f'_{-s}\Omega)^{b_1\dots b_m}(x).
\label{RR}
\ee
The expression of $\f'_{-s}\vec{\O}|_x$ can be worked out from 
$\f'_s\vec{\O}|_{\f_s(x)}=\beta_s(\f_s(x))\vec{\O}|_{\f_s(x)}$ 
and it turns out to be $\f'_{-s}\vec{\O}|_x=
\beta^{-1}_{s}(x)\vec{\O}|_x$.
On the other hand, since the null vectors $\k^{j}|_x$ forming the $m$-form are 
eigenvectors of $\f^{*}_s\G|_x$ with the same eigenvalue $\l_s$ 
for $s>0$ we have the property 
$(\f^{*}_s\rmg)_{ab}(x)k^{j\ b}(x)=\l_s(x)k^j_{\ a}(x)$. 
The insertion of both 
results in (\ref{RR}) leads to (\ref{omega}) at once.\N

In most of our calculations we will use the equation involving $\O$  
rather than its vector counterpart. In the particular cases of $\mu_{\xiv}$ 
consisting of one or two linearly independent null vectors lemma \ref{form1} 
can be written as $\f'_s\k|_{\f_s(x)}\propto\k|_{\f_s(x)}$ with $\k$ any element 
of $\mu_{\xiv}$. 

An important rank-2 tensor related with the $m$-form $\O$ introduced in 
proposition \ref{form2} is its Lorentz tensor $T\{\O\}=\S$ which, as explained 
in appendix A, is an element of $\DP_2^+(V)$ which can be normalized as
$S_{ac}S^{c}_{\ b}=\rmg_{ab}$ by choosing $\O$ such that 
$\O\cdot\O=\Omega_{a_1\dots a_m}\Omega^{a_1\dots a_m}=(-1)^{m-1}2m!$ 
(unless $\O\cdot\O=0$ in which case 
$S_{ac}S^{c}_{\ b}=0\Leftrightarrow \S=f\K\otimes\K$, $\K$ null;
in our setting this only happens in the degenerate case). 
The only possible eigenvalues for the normalized tensor $\S$ arising 
in the nondegenerate case are +1 and -1.

We can now write down equations for $\f^{*}_s\G$ when we have a one-parameter
submonoid of causal symmetries. The very first thing we know is that 
$\f^{*}_s\G\in\DP^+_2(V)$ so we can apply theorem \ref{DECOMP} to this tensor
and write it as 
$$
\f^{*}_s\G=\alpha_s\G+\U_s,\ \ \alpha_s>0,
$$ 
where $\U_s\in\DP^+_2$ gathers the terms of the 
decomposition which are not proportional to the metric tensor. The term 
proportional to the metric always appears if $s$ is small enough due to the 
continuity in the parameter  $s$ (actually it can be shown by algebraic arguments
that $\alpha_s\neq 0$, $\forall s\in\r^+\cap I$) whence 
$\mu(\f^{*}_s\G)=\mu(\U_s)=\mu_{\xiv}$, $\forall s\in I\cap\r^+$. 
In the case of more than one linearly independent canonical null direction, a new application of theorem 
\ref{DECOMP} to $\U_s$ tells us that $\S$ will always be 
the first term of the decomposition so it would be useful to know how it behaves 
under the pull-back of $\f_s$. This calculation is easily done using equation 
(\ref{s-e-f}) with $\Sg=\O$ which yields the formula for the nondegenerate case
with $\chi(\mu_{\xiv})=m>1$:
\bnr
\fl(\f^{*}_sS)_{ab}(x)=\fr{(-1)^{m-1}}{(m-1)!}
\left[(\f^{*}_s\Omega)_{aa_2\dots a_m}(x)(\f^{*}_s\rmg)_{ba_1}(x)
(\f'_{-s}\Omega)^{a_1a_2\dots a_m}(x)-\right.\\
-\left.\fr{1}{2m}(\f^{*}_s\rmg)_{ab}(x)
(\f^{*}_s\Omega)_{a_1\dots a_m}(x)(\f'_{-s}\Omega)^{a_1\dots a_m}(x)\right],
\enr
where everything is known. Moreover is quite simple 
to realize that the normalization imposed on $\O$ in this case
 allows us to rewrite its 
invariance law as $\f'_{-s}\vec{\O}|_x=\vec{\O}|_x/\s_s(x)$ by just pulling 
back the normalization condition.
Plugging $\f^{*}_s\O|_x$ and $\f'_{-s}\vec{\O}|_x$ into the previous equation 
we get:
\bnr
\fl(\f^{*}_sS)_{ab}=\fr{(-1)^{m-1}}{(m-1)!}
\left[\Omega_{aa_2\dots a_m}(\f^{*}_s\rmg)_{ba_1}\Omega^{a_1a_2\dots a_m}-
\fr{1}{2m}(\f^{*}_s\rmg)_{ab}\, \O\cdot\O\right].
\enr 
Replacing $\f^{*}_s\G$ by $\alpha_s\G+\U_s$ and rearranging terms we obtain the 
final formula
\be
(\f^{*}_sS)_{ab}=\alpha_sS_{ab}+(U_s)_{bc}S^{c}_{\ a},\ \ \chi(\mu_{\xiv})>1,
\label{s-dS}
\ee
from where the property $(U_s)_{bc}S^c_{\ a}=(U_s)_{ac}S^c_{\ b}$ 
follows due to the symmetry of $(\f^*_sS)_{ab}$.

The degenerate case is even simpler as $\S=\K\otimes\K$ 
(we set $f=1$) and 
we know that $\f^{*}_s\K|_x=\g_s(x)\K|_x$ for some function $\g_s(x)$. 
Whence the
transformation law for $\S=\K\otimes\K$ in the degenerate case is simply
\be
\f^{*}_s\S|_x=\g^2_s(x) \S|_x,\ \ \ \ \
\chi(\mu_{\xiv})=1.
\label{degenerate}
\ee
If there are no canonical null directions then things are far more involved
because we cannot prove by the same means the existence 
of a timelike future-directed vector field invariant under the 
submonoid $\{\f_s\}_{s\in I}$ of causal symmetries. Even if such a field 
$\u$ did exist, the invariance property $\f'_s(\u|_x)\propto\u|_{\f_s(x)}$, 
$s>0$ would not imply that $\u|_x$ is an eigenvector of $(\f^*_s\G)|_x$ 
(see Example 1 below) as happened in the cases with $\chi(\mu_{\xiv})=1,2$. 
Hence, in this last case we will have to deal with 
two different situations according to whether $\u|_x$ is or not an eigenvector 
of $(\f^*_s\G)|_x$.

\medskip
\noindent
{\bf Example 1.} Let $(V,\G)$=$(\L,\bfeta)$ be flat Minkowski spacetime 
and consider any linear automorphism $\hat{\T}$ of $\L$ seen as a vector space.
Then $\hat{\T}$ defines in an obvious way a transformation $\f\in Diff(\L)$ 
which, if the time orientation is preserved, will be in $\C(\L,\bfeta)$ 
whenever 
$(\f^*\bfeta)_{ab}=\hat{T}^p_{\ a}\eta_{pq}\hat{T}^q_{\ b}\in\DP^+_2(\L)$. 
Clearly it is possible to define one-parameter
monoids of this type of transformations such that 
$\{\f_s\}_{s>0}\in\C(\L,\bfeta)$ 
and having the properties described in this section. An additional property 
holding in this case is the existence of an invariant causal direction for each 
value of $s$ (see theorem \ref{fixedvector} of appendix B), so if we denote 
by $\u_s$ such direction we get the property 
$\f'_{s}\u_s=\hat{\T}_{s}(\u_s)\propto\u_s$. 
Nevertheless, in the case that $\{\f_s\}_{s>0}$ has no canonical null 
directions, the (necessarily) timelike vector $\u_s$ will not be in general 
an eigenvector of $\f_{s}^*\bfeta$ unless 
$(u_{s})_{a}(\hat{T}_{s})^a_{\ b}\propto (u_{s})_{b}$.     

In the next section we will finally derive the differential 
conditions we are seeking. They are the infinitesimal counterpart of the 
equations of this section and we shall discuss some of their properties.                        

\section{Infinitesimal theory}
\label{teoria-infinitesimal}
\begin{defi}
A smooth vector field $\xiv$ defined on an entire Lorentzian manifold is 
said to be {\em causal preserving} if the one-parameter group $\{\f_s\}_{s\in I}$ 
generated by $\xiv$ is such that $\{\f_s\}_{s\in I\cap\r^{+}}\in\C(V,\G)$. 
A casual-preserving vector field is called (non-) degenerate if the corresponding 
$\{\f_s\}_{s\in I\cap\r^+}$ is (non-) degenerate.
\label{causal-symmetry}
\end{defi}
\noindent
{\bf Remark.} This definition is rather general in the sense that 
$\{\f_s\}_{s\in\r^+\cap I}$ may give rise to different components $\Um$ and $\Nm$
of $V$ with $\chi(\mu_{\xiv}|_x)$ changing from one set to another. Proposition
\ref{invariance} allows us to study only cases in which $V$ consists of
just one of these sets and so all the causal-preserving vector fields 
(CPVF from now on) of this paper will 
be understood in this way.

As stated in the previous section, our aim is to find differential 
conditions for the causal symmetries analogous to those of other known
symmetries. We need a lemma.
\begin{lem}
Let $\{\T_s\}_{s\in I}$ be a one-parameter family of rank-r covariant 
(contravariant) tensors 
differentiable in the parameter $s$ such that $\T_{s_0}={\bf 0}$ and assume 
further that 
$\T_s\in\DP^+_r(V)$ for $s\in[s_0,s_0+\epsilon)$ . Then 
$d\T_s/ds|_{s=s_0}\equiv\dot{\T}_ {s_0}\in\DP^+_r(V)$ 
(or its contravariant counterpart).     
\label{DP-continuation}
\end{lem}
\P Define the family of functions 
$f_{\u_1,\dots,\u_r}(s)=\T_s(\u_1,\dots,\u_r)$ for each
$r$-tuple of causal future-directed vectors $\u_1,\dots,\u_r$. All these functions 
vanish at $s=s_0$ and are non-negative if $s$ lies in $[s_0,s_0+\epsilon)$ due to the 
hypotheses of the lemma so they are nondecreasing at $s=s_0$ as functions of the 
parameter $s$ which entails $d\T_s/ds|_{s=s_0}(\u_1,\dots,\u_r)\equiv
f^{'}_{\u_1,\dots,\u_r}(s_0)=\dot{\T}_{s_0}(\u_1,\dots,\u_r)\geq 0$ for every set
of causal future-directed vectors $\u_1,\dots,\u_r$.\N     

The next proposition characterizes $\mu_{\xiv}$ as the set of null vectors
onto which $\lie\G$ vanishes so that this set depends only on the CPVF
$\xiv$.
\begin{prop} $\k\in\mu_{\xiv}|_x$ if and only if $(\lie\G)|_x(\k,\k)=0$.
\label{MU}
\end{prop}
\P The ``only if'' implication comes easily
from the property $\f^*_s\G|_x(\k,\k)=0$ holding 
for every $\k\in\mu_{\xiv}$. To prove the ``if'' assertion 
we proceed by contradiction: let us assume that a null
vector $\k|_x$ exists such that $\f'_s\k|_{\f_s(x)}$ is 
timelike for 
 $s\in\r^+\cap I$ and $\lie\G|_x(\k,\k)=0$. Lemma 2.5 in PI or a direct 
calculation shows that 
$\f'_s\k|_{\f_s(x)}=c_s\k|_{\f_s(x)}+\n_s|_{\f_s(x)}$ for some $c_s>0$ and 
$\n_s|_{\f_s(x)}$ null and future directed. Therefore 
$\n_s|_x=\f'_s\k|_x-c_s\k|_x$ is a one-parameter family of causal 
vectors which comply with the hypotheses of lemma \ref{DP-continuation} 
for $s_0=0$ so $\dot{\n}_0|_x=-\lie\k|_x-\dot{c}_0\k|_x\in\Z^+_{x}$. 
On the other hand if we apply
the Lie derivative to $k^a\rmg_{ab}k^b$ and use  
$0=(\lie\rmg)_{ab}k^ak^b$ we get $k_a\lie k^a=0$ so $\dot{\n}_0|_x$ is causal and 
orthogonal to $\k|_x$ which is only possible if both vectors are null 
and proportional to each other, that is to say, $\lie\k|_x\propto\k|_x$ which 
integrated yields 
$\f'_s\k|_{\f_s(x)}\propto\k|_{\f_s(x)}$ against the assumption of 
$\f'_s\k$ being timelike.\N  

These two results and the calculations of the last part of the previous section is 
all what is needed to establish one of the main results of this paper. 
\begin{theo}
Let $\{\f_s\}_{s\in I\cap\r^+}$ be a one-parameter submonoid of causal symmetries
generated by the vector field $\xiv$. Then $\lie\G$ and $\lie\S$ adopt
the expressions 
written below, according to the number of linearly independent elements of 
$\mu_{\xiv}$:
\begin{enumerate}
\item if $\chi(\mu_{\xiv})>1$ then:
\bea
\lie\rmg_{ab}=\alpha\rmg_{ab}+\beta S_{ab}+Q_{ab},\label{primera-a}\\ 
\lie S_{ab}=\beta\rmg_{ab}+\alpha S_{ab}+Q_{ac}S^{c}_{\ b},
\label{primera-b}
\eea
\item if $\chi(\mu_{\xiv})=1$ (degenerate CPVFs) then: 
\be
\fl\lie\rmg_{ab}=\alpha\rmg_{ab}+\beta k_ak_b+Q_{ab},\ \   
\lie S_{ab}=2\g S_{ab}\Leftrightarrow \lie k_a=\g k_a,
\label{degenerada}
\ee
with $S_{ab}=k_ak_b$. 
\item and, if there are no canonical null directions:
\be
\lie\rmg_{ab}=\alpha\rmg_{ab}+U_{ab}.
\label{empty}
\ee 
\end{enumerate}
In all cases $\U$, $\Q\in\DP^+_2(V)$, $\mu(\Q)\supset\mu_{\xiv}$ and $\beta>0$.
\label{differential}
\end{theo} 
\P The family $\{\U_s\}_{s\in I}$ introduced in previous section behaves in the 
way of lemma \ref{DP-continuation} with $s_0=0$ so the formula for $\f^{*}_s\G$ 
in terms of $\U_s$ yields the equation:
\be
\left.\fr{d\f^{*}_s\G}{ds}\right|_{s=0}=\lie\G=\alpha\G+\U,\ 
\alpha=\left.\fr{d\alpha_s}{ds}\right|_{s=0},\ \ 
\U=\left.\fr{d\U_s}{ds}\right|_{s=0}.
\label{primera}
\ee
In the case of $\chi(\mu_{\xiv})>1$, this equation must be supplemented 
with the derivative with respect to $s$ of (\ref{s-dS}):
\be
\left.\fr{d\f^{*}_s S_{ab}}{ds}\right|_{s=0}=
\lie S_{ab}=\alpha S_{ab}+(U)_{ac}S^{c}_{\ b},
\label{segunda}
\ee
where the tensor $\U\in\DP^{+}_2(V)$. Equation (\ref{primera}) and 
proposition \ref{MU} imply that $\mu(\U)=\mu_{\xiv}$ so if we apply
theorem \ref{DECOMP} to $\U$ we may write it as $\U=\beta\S+\Q$ 
where $\beta$ must be strictly positive since it is the coefficient of the first 
term of the decomposition and $\Q$ is a causal tensor collecting the 
rest of the terms, and thus $\mu_{\xiv}=\mu(\U)\subset\mu(\Q)$. 
Inserting this new expression of $\U$ into 
equations (\ref{primera}) and (\ref{segunda}) yields (\ref{primera-a}) and 
(\ref{primera-b}). In the degenerate case $\U$ 
has only a single null direction so that $\S\propto \K\otimes\K$.  
The second equation of (\ref{degenerada}) is the derivative of 
(\ref{degenerate}) at $s=0$. Finally,
in the case with no canonical null directions we are only left with equation 
(\ref{primera}) where now $\U$ is a causal tensor with no principal directions.\N 

\medskip
\noindent
{\bf Remarks.}
\begin{enumerate}
\item  As we explained in section \ref{causal-symmetries} the set 
$\C(V,\G)$ is invariant under conformal rescalings of the metric tensor $\G$.
Therefore, if $\{\f_{s}\}_{s\in\r^+\cap I}$ is a submonoid of causal symmetries 
with respect to $\C(V,\G)$ it will also be so for any metric $\sigma\G$, 
 $\sigma(x)>0$ $\forall x\in V$ (with the same number of linearly independent 
  canonical null directions as is obvious). As a result 
of this conformal invariance the function $\alpha(x)$ which appears in  
equations (\ref{primera-a}-\ref{empty}) is 
redefined by conformal rescaling of the metric $\G\rightarrow\sigma\G$ as
 $\alpha\rightarrow\sigma^{-1}\lie\sigma+\alpha$ whereas $\Q$, $\U$ and $\S$ 
are just rescaled by $\sigma$. Moreover, we can set 
$\sigma^{-1}\lie\sigma+\alpha$ to zero by choosing an appropriate 
initial condition for $\s$. This choice can only be made globally if the vector 
$\xiv$ is complete.                                                                             

\item The remainder $\Q$ will depend on each CPVF although some general 
properties are common to every case.  Given that $\mu_{\xiv}\subset\mu(\Q)$ if 
$\mu_{\xiv}\neq\emptyset$, every element of 
$\mu_{\xiv}$ is a null eigenvector of $\Q$ which implies that 
$$
Q_{a}^{\ b}\Omega_{ba_2\dots a_m}= 
\l\Omega_{aa_2\dots a_m}\Rightarrow Q_{ab}S^b_{\ c}=Q_{cb}S^b_{\ a},
$$
where $\l$ is the associated eigenvalue. In the case of $\mu_{\xiv}$ being empty,
it is still possible to get an equation similar to (\ref{primera-a}) as follows: 
since the tensor $U_{ab}$  of equation (\ref{empty}) has no principal directions,
the first term of its canonical  decomposition $\beta S_{ab}$, $\beta>0$, is now 
proportional to the Lorentz tensor $S_{ab}$ associated to the timelike 
eigenvector field $\u$ of $\lie\G$ (we  also keep the notation $Q_{ab}$ for the 
remainder appearing in this decomposition). Assuming 
the normalization $u_au^a=1$, this tensor reads $S_{ab}=2u_au_b-\rmg_{ab}$.
As we commented in example 1, invariance properties under the one-parameter 
submonoid $\{\f_s\}_{s\in\r^+\cap I}$ are in general unknown in this case so, 
as far as we know, 
equation (\ref{primera-b}) will not be true in general. Nonetheless,
in the particular case of having a timelike direction invariant under 
$\{\f_s\}_{s\in\r^+\cap I}$, say $\v$, we can still
use part of the calculations above. For instance
the Lie derivative of the 1-form $v_a$ is given by:
\bea
\lie v_a=\lie(\rmg_{ap}v^p)=(\alpha\rmg_{ap}+\beta S_{ap}+Q_{ap})v^p+\g v_a,
\label{v}
\eea
where we have used $\lie v^a=\g v^a$ for some function $\g(x)$.
The case with $\v=\u$ will be called {\em aligned} and has the same infinitesimal 
treatment as the others with $\chi(\mu_{\xiv})>1$ because, as follows from 
the previous equation, $\lie u_{a}\propto u_a$ and a very easy calculation 
leads then to (\ref{primera-b}) with the new Lorentz tensor $S_{ab}$. Therefore 
all the results arising from theorem \ref{differential} valid for the case 
with $\chi(\mu_{\xiv})>1$ will also hold for the aligned case with the previous 
definition of $\S$ in terms of $\u=\v$. 
  
\item Another important point is that $P_{ab}=(g_{ab}+S_{ab})/2$ is a projector 
onto the subspace $Span\{\mu(\S)\}=\mu_{\xiv}$ if $\S$ is nondegenerate as 
it is explicitly shown in proposition \ref{projection}. This applies 
to case ({\it i}) of previous theorem, and also to the aligned case 
(with $\mu_{\xiv}=\emptyset$) where the subspace is now $Span\{\u\}$.
Since $Span\{\mu(\S)\}$ (or $Span\{\u\}$) is an invariant subspace of $Q_{ab}$, 
this tensor must be invariant under the
action of the projector which results in
$$
Q_{a}^{\ c}(\rmg_{cb}+S_{cb})=\l(\rmg_{ab}+S_{ab}).
$$
With the aid of this result
we can show explicitly the partly conformal character of these transformations by 
adding 
equations (\ref{primera-a}) and (\ref{primera-b}) to get: 
\be
\fl\lie P_{ab}=(\alpha+\beta)
\fr{1}{2}\left(\rmg_{ab}+S_{ab}\right)+
Q_{ac}\fr{1}{2}\left(\delta^{c}_{\ b}+S^{c}_{\ b}\right)=
(\alpha+\beta+\l)P_{ab}.
\label{partly}
\ee
We see that these transformations can be regarded as conformal symmetries of the 
projector $P_{ab}$ i.e. they are mapping conformally $Span\{\mu_{\xiv}|_x\}$
($Span\{\u|_{x}\}$) onto $Span\{\mu_{\xiv}|_{\f_s(x)}\}$ 
($Span\{\u|_{\f_s(x)}\}$) for all $s\in I$ and all $x\in V$. 
\end{enumerate}

\noindent
{\bf Example 2.} As an example of the foregoing, take as $(V,\G)$ the four 
dimensional line element given by
\be
ds^2=(1+B^2\rho^2)^2(dt^2-d\rho^2-dz^2)-(1+B^2\rho^2)^{-2}\rho^2d\f^2
\label{MELVIN},
\ee
where the coordinate ranges are 
$0<\rho<\infty,\ -\infty<t,z<\infty,\ \ 0<\phi<2\pi$. This is a solution of 
Einstein-Maxwell equations with a uniform magnetic field $2B$ along the $z$-axis
 known as Melvin solution \cite{MELVIN,BONNOR,KRAMERS}.
Consider the one-parameter group of diffeomorphisms 
$\f_s:\rho\rightarrow \rho+s$ whose generator is the spacelike vector
field $\xiv=\fr{\d}{\d\rho}$. The pull-back of the metric tensor under this 
group turns out to be:
\bnr
\hspace{-1cm}\f^*_s\G=(1+B^2(\rho+s)^2)^2(dt^2-d\rho^2-dz^2)-
\fr{(\rho+s)^2}{(1+B^2(\rho+s)^2)^2}d\phi^2=\\
\hspace{-1cm}\fr{(1+B^2(\rho+s)^2)^2}{(1+B^2\rho^2)^2}((\z^0)^2-(\z^1)^2-(\z^2)^2)-
\fr{(\rho+s)^2(1+B^2\rho^2)^2}{\rho^2(1+B^2(\rho+s)^2)^2}(\z^3)^2,
\enr
where we have written $\f^*_s\G$ in the cobasis of orthonormal 1-forms
$\z^a$ (dual to its orthonormal basis of eigenvectors). According to proposition 
\ref{DP-condition}, $\f^{*}_s\G\in\DP^+_2(V)$ if 
the inequality
$$
\fr{(1+B^2(\rho+s)^2)^2}{\rho+s}\geq\fr{(1+B^2\rho^2)^2}{\rho}
$$
is fulfilled. It is fairly simple to check 
that this happens if $\rho>1/\sqrt{3B}$
 and $s>0$ so $\xiv$ is a CPVF in the region with
$\rho>1/\sqrt{3B}$ whereas $-\xiv$ is so in $\rho<1/\sqrt{3B}$. From the 
formula of $\f^*_s\G$ one gets
$Span\{\mu_{\xiv}\}=Span\{\partial_{t},\partial_{\rho},\partial_{z}\}$. 

It is also an easy task to show that $\lie\G$ adopts the form displayed 
in theorem \ref{differential} by just rearranging it a little bit:    
\bnr
\fl\lie\G=4B^2\rho(1+B^2\rho^2)(dt^2-d\rho^2-dz^2)-
\fr{2\rho(1-B^2\rho^2)}{(1+B^2\rho^2)^3}d\phi^2=\alpha\G+\U,\\
\fl\U=\fr{3B^2\rho^2-1}{\rho(1+B^2\rho^2)}\S,
\ \ \ \S=(\z^0)^2-(\z^1)^2-(\z^2)^2+(\z^3)^2,\ \ \alpha=\fr{1}{\rho}
\enr
which means that $\lie\G$ is a linear combination of $\G$ and the 
Lorentz tensor $\S$. This is one of the defining features 
of {\it nondegenerate  bi-conformal vector fields} \cite{PIV} which, 
in particular, are the only possible CPVF if
$\chi(\mu_{\xiv})$ is the spacetime dimension minus one (see next proposition).   
\begin{prop}
Every CPVF with $n-1$ linearly independent canonical null directions
for $n\geq 3$ satisfies
$$
\lie\rmg_{ab}=\alpha\rmg_{ab}+\beta S_{ab},\ \ \lie S_{ab}=\alpha S_{ab}+\beta\rmg_{ab},
$$
where $\S$ is the Lorentz tensor built up from the set of linearly independent  
canonical null directions and $\beta>0$.
\label{n-1}
\end{prop}
\P  Since the number of linearly independent canonical null directions is greater or equal than 2 we know
that $\xiv$ will satisfy equations (\ref{primera-a}) and 
(\ref{primera-b}) with $\mu_{\xiv}=\mu(\S)$ so $\S$ is the Lorentz tensor of 
a simple $n-1$ form. Therefore the tensor $\Q$ appearing in those equations is a 
causal tensor with $n$ linearly independent principal directions which means 
that it is proportional to the metric tensor. Putting $\Q=\g\G$ in 
(\ref{primera-a}) and (\ref{primera-b}) we get:
$$
\lie\rmg_{ab}=\alpha\rmg_{ab}+\beta S_{ab}+\g\rmg_{ab},\ \ \ 
\lie S_{ab}=\alpha S_{ab}+\beta\rmg_{ab}+\g S_{ab}
$$
which is the desired form if we relabel $\alpha\rightarrow\alpha+\g$.
\N

Nondegenerate bi-conformal vector fields are the main subject of \cite{PIV} where 
a thorough study of the properties and applications of these symmetries is 
performed.
  
Equation (\ref{primera-a}) can always be written in terms of the Lie derivative 
of the contravariant metric $\rmg^{ab}$ by using the formula 
$\lie\rmg^{ab}=-\rmg^{ap}\rmg^{bq}\lie\rmg_{pq}$:
\be
\lie\rmg^{ab}=-\alpha\rmg^{ab}-\beta S^{ab}-Q^{ab}     
\label{contravariant}
\ee
which is valid regardless of the cases (i)-(iii) of theorem \ref{differential}. 
Then, formula (\ref{primera-b}) in theorem \ref{differential} turns out to admit 
more equivalent forms as we prove next.  
\begin{prop}
If equation (\ref{primera-a}) of point {\it (i)} in theorem \ref{differential}
 holds, then equation (\ref{primera-b}) is equivalent to either of the following: 
\be
\fl\lie\vec{\O}=\fr{-m}{2}(\alpha+\beta+\l)\vec{\O}, \ \ 
\lie\O=\fr{m}{2}(\alpha+\beta+\l)\O,\ \ \lie S^a_{\ b}=0,
\label{forms-invariance}
\ee
where the $m$-form $\O$ gives rise to the Lorentz tensor 
$\S$ according to the conventions introduced at the end of section 
\ref{Continuous} (in particular 
$m=\chi(\mu_{\xiv})=${\em dim}$(Span\{\mu(\S)\})>1$).
\label{forms}
\end{prop}
\P Let us first assume that equations (\ref{primera-a}) and (\ref{primera-b})
are fulfilled. Then
 equation (\ref{contravariant}) 
can be used to work out $\lie S^a_{\ b}$ which turns out to be:
\bnr
\fl\lie S^{a}_{\ b}=\lie(\rmg^{ap}S_{pb})=(-\alpha\rmg^{ap}-\beta S^{ap}-Q^{ap})S_{pb}+
\rmg^{ap}(\alpha S_{pb}+\beta\rmg_{pb}+Q_{pc}S^{c}_{\ b})=0
\enr
where the property $Q_{ap}S^{p}_{\ b}=Q_{bp}S^{p}_{\ a}$ has been used.
By means of the invariance
of $S^{a}_{\ b}$ we can obtain the sought expressions for the 
Lie derivatives of the 
normalized forms $\vec{\O}$ and $\O$ as follows: from the algebraic properties 
of $\S$ any $\k\in\mu_{\xiv}$ is a null
eigenvector of $S^a_{\ b}$ with unit eigenvalue i.e. $S^{a}_{\ b}k^b=k^a$.
Taking the Lie derivative of this equation we deduce at once that 
$S^a_{\ b}\lie k^b=\lie k^a$ which entails $\lie\k\in Span\{\mu_{\xiv}\}$. 
Therefore $\lie\vec{\O}=\lie(\k^1\wedge\dots\wedge\k^m)=
\psi\k^1\wedge\dots\wedge\k^m$ for 
some proportionality factor to be determined later. Using this, we can work out
$\lie\O$ in the usual fashion getting
\be
\lie\O=[\psi+m(\alpha+\beta+\l)]\O,
\ee
where the expression for $\lie\G$ has been used. The function
$\psi$ is fixed by Lie derivating the normalization condition 
$\O\cdot\O=2m!(-1)^{m-1}$ yielding $\psi=\fr{-m}{2}(\alpha+\beta+\l)$.

Suppose now that either of the equations of (\ref{forms-invariance}) 
together with (\ref{primera-a}) hold. Equation (\ref{primera-b}) clearly comes 
from $\lie S^a_{\ b}=0$ so it is enough to prove the equivalence of all 
the equations in (\ref{forms-invariance}) if
(\ref{primera-a}) holds. In the previous paragraph we already proved that 
$\lie S^a_{\ b}$ implies 
the formulae involving $\lie\O$ and $\lie\vec{\O}$ and clearly these 
two equations are equivalent if (\ref{primera-a}) is true. 
Therefore we only need to show that the first couple of 
(\ref{forms-invariance}) implies $\lie S^a_{\ b}=0$ which again is 
straightforward using (\ref{primera-a}) 
(and its consequence (\ref{contravariant})).\N  

\medskip
\noindent
{\bf Remark.} 
Equations (\ref{forms-invariance}) are nothing but the infinitesimal formulation
of lemma \ref{form1} and proposition \ref{form2} and they 
clearly state the conformal invariance of $\vec{\O}$ and $\O$ as one of the 
key properties of causal symmetries with a number of linearly independent 
canonical null directions greater than one. For the case without
canonical null directions but {\em aligned}, so that there is an invariant 
timelike direction $\u$, Eqs.(\ref{forms-invariance}) still hold 
with $\O =\vu$ and $\S\propto\T\{\vu\}$ the corresponding Lorentz tensor.
They are thus 
equivalent to $\lie S_{ab}=\alpha S_{ab}+\beta\rmg_{ab}+Q_{ac}S^c_{\ b}$ for 
this Lorentz tensor; see remark {\it (ii)} of theorem \ref{differential}.

Theorem \ref{differential} involves a stability property under
repeated action of the differential operator $\lie$ in a sense which we 
are going to explain next. The set of principal directions of the tensor $\U$ 
coincides with the set of canonical null directions of the submonoid so one 
might expect that $\U$ satisfied a differential condition similar to 
that fulfilled by $\G$. To prove this, we need a preparatory lemma.
\begin{lem}
Let $\T\in\DP_2^+(V)$ be a future tensor such that 
$\mu(\T|_x)\supseteq\mu_{\xiv}|_x$ $\forall x\in V$ and 
$\chi(\mu_{\xiv})\neq 1$. Then there exists a smooth function $\psi$ such that 
$\lie\T-\psi\T\in\DP_2^+(V)$ and $\mu(\lie\T-\psi\T)\supseteq\mu_{\xiv}$.
\label{stability}
\end{lem}
\P The assumption $\mu(\T|_x)\supseteq\mu_{\xiv}|_x$ implies, according to 
proposition \ref{S=M}, that $\mu(\f^{*}_s\T|_x)=\s(\f^*_s\T|_x)=\mu_{\xiv}|_x$ 
$\forall x\in V$, $\forall s>0$. 
As $\chi(\mu_{\xiv})\neq 1$ we can apply lemma \ref{canonical} to the 
future tensors $\f^{*}_{s_1}\T$ and $\f^{*}_{s_2}\T$
and write for every $s_2,s_1\geq 0$:
$$
\f^{*}_{s_2}\T=\alpha_{s_2,s_1}\f^{*}_{s_1}\T+\R_{s_2,s_1},
$$
where $\R_{s_2,s_1}\in\DP^+_2(V)$ and $\alpha_{s_2,s_1}$ is a positive function 
which can always be chosen in such a way that $\alpha_{s_1,s_1}=1$. This choice 
implies that $\R_{s_1,s_1}=0$ so if we let $s_1$ fixed we can construct the 
one-parameter family $\T_s\equiv\f^{*}_s\T-\alpha_{s,s_1}\f^{*}_{s_1}\T$ 
fulfilling the requirements of lemma \ref{DP-continuation} with $s_0=s_1$ from 
what we conclude that 
$$
\left(\left.\fr{d(\f^{*}_s\T)}{ds}\right|_{s=s_1}-
\left.\fr{d\alpha_{s,s_1}}{ds}\right|_{s=s_1}(\f^{*}_{s_1}\T)\right)\in\DP^+_2(V), 
$$ 
where $s_1\in[0,\epsilon)$ (note that this tensor vanishes when acting over any 
element of $\mu_{\xiv}$).
By putting $s_1=0$ in this last equation both results follow.\N

\noindent
{\bf Remark.} Note that this lemma holds even if $\mu_{\xiv}=\emptyset$ 
\begin{theo}
For a nondegenerate CPVF $\xiv$ we can find smooth functions 
$\alpha$, $\alpha_1\dots,\alpha_r,\dots$ (free functions) such that 
$$
\Q_{r+1}\equiv(\lie-\alpha_r)\cdots(\lie-\alpha_1)(\lie-\alpha)\G\in\DP^+_2(V),\ 
\forall r\in\NAT.
$$  
Furthermore, $\mu(\Q_r)\supseteq\mu_{\xiv}$, $\forall r\in\NAT$.
\label{stable-symmetry}
\end{theo}
\P The causal tensor $\U$ has $\mu_{\xiv}$ as its set of principal  
directions and by assumption $\chi(\mu_{\xiv})\neq 1$ so we can apply lemma 
\ref{stability} to deduce the existence of a causal 
tensor $\Q_2$ and a function $\alpha_1$ such that 
$\Q_2\equiv\lie\U-\alpha_1\U$. Clearly $\Q_2$ complies again with 
lemma \ref{stability} so we get in the same fashion a causal tensor $\Q_3$ and 
a function $\alpha_2$ such that $\Q_3=\lie\Q_2-\alpha_2\Q_2$ and so on. In this 
way we obtain  a chain of equations
\be
\eqalign{
\lie\G-\alpha\G=\U\nonumber\\
\lie\U-\alpha_1\U=\Q_2\nonumber\\
\lie\Q_2-\alpha_2\Q_2=\Q_3\nonumber\\
\cdot\cdot\cdot\cdot\cdot\cdot\cdot\cdot\cdot\cdot\cdot\cdot\cdot\cdot\cdot\cdot
\nonumber\\
\lie\Q_r-\alpha_r\Q_r=\Q_{r+1}\nonumber\\
\cdot\cdot\cdot\cdot\cdot\cdot\cdot\cdot\cdot\cdot\cdot\cdot\cdot\cdot\cdot\cdot,}
\label{chain}
\ee
where the tensors on the right hand side of each equation belong to $\DP^+_2(V)$
and contain $\mu_{\xiv}$ as a subset of its set of principal directions 
due to lemma \ref{stability}. The theorem follows because
$\Q_r=(\lie-\alpha_{r-1})(\lie-\alpha_{r-2})\cdots(\lie-\alpha_1)
(\lie-\alpha)\G$.\N

The chain of equations (\ref{chain}) used in the proof of this theorem
can be reduced to a finite number by noticing that we
must reach an index $j$ for which the causal tensor $\Q_{j+1}$ can be written in
terms of the previous causal tensors as 
$\Q_{j+1}|_x=\phi_0(x)\G|_x+\phi_1(x)\U|_x+\sum_{k=2}^{j}\phi_k(x)\Q_{k}|_x$ 
for some smooth 
functions $\phi_0(x),\phi_1(x),\phi_2(x),\dots,\phi_j(x)$. Thus the last equation of the chain
can be taken to be, by redefining $\alpha_j$ if necessary
\be
\lie\Q_j-\alpha_j\Q_j=\phi_0\G+\phi_1\U+\sum_{k=2}^{j-1}\phi_k\Q_{k}.
\label{last}
\ee

Up to now we have provided a thorough study of the properties of CPVFs 
and we have found the necessary conditions required for a vector field 
to be causal preserving. However, we still have no means to decide whether a 
given vector field is causal preserving or not. Next, we solve this 
problem and show that the 
necessary conditions of theorem \ref{differential} are in fact 
sufficient by proving its converse.
\begin{theo}
Let $\xiv$ be a smooth complete vector field and assume that 
there exists a function $\alpha$ such that 
$\R\equiv\lie\G-\alpha\G\in\DP^+_2(V)$ with $\R$ keeping the same 
number of principal null directions on $V$. 
\begin{itemize}
    \item If $\mu(\R)=\emptyset$ then $\xiv$ is a CPVF with 
    $\mu_{\xiv}=\emptyset$. 
    \item If $\mu(\R)\neq \emptyset$ and $\lie\O\propto\O$ where $\O$ 
    is a non-zero $m$-form of maximum degree on $Span\{\mu(\R)\}$,  
    then $\xiv$ is a CPVF with $\mu_{\xiv}=\mu(\R)$.
\end{itemize}
\label{converse}
\end{theo}
\P We will perform the proof separating out the cases with 
$\mu(\R)\neq\emptyset$ and $\mu(\R)=\emptyset$. Assume first that the causal 
tensor $\R$ has no principal directions. This means that, for every null vector 
$\k$, $\lie\G|_x(\k,\k)>0$ $\forall x\in V$ so that the function 
$f_{\k}(s)\equiv\f^*_s\G|_x(\k,\k)$ is positive for all 
$s\in(0,\epsilon)$ and negative in $(-\epsilon',0)$ for some
small $\epsilon,\epsilon'>0$. In fact, the result holds for all 
positive values of $\epsilon',\epsilon$ because, if it did not, 
there would be a first value $s_0>0$ (the proof for negative values of $s$ is 
analogous) such that 
$0=f_{\k}(s_{0})=\G|_{\f_{s_0}(x)}(\f'_{s_0}\k,\f'_{s_0}\k)$ 
implying that $\f'_{s_0}\k$ is a null vector so that, using our 
assumption again, $\lie\G|_{\f_{s_0}(x)}(\f'_{s_0}\k,\f'_{s_0}\k)$ 
should also be strictly positive.  Thus we could write 
$0<\lie\G|_{\f_{s_0}(x)}(\f'_{s_0}\k,\f'_{s_0}\k)=
(\f^*_{s_0}\lie\G)|_x(\k,\k)=d\f^*_s\G|_x/ds|_{s=s_0}(\k,\k)$ 
where in the last step the standard formula
\be
\fr{d\f^{*}_s\T}{ds}=\f^{*}_s\lie\T,
\label{lie-property}
\ee
must be used. Hence, $f_{\k}(s)$ would be strictly increasing at $s=s_0$ 
too, which is obviously a contradiction.

If $\mu(\R)\neq\emptyset$ then this set consists of the 
future-pointing null vector fields $\k\in T(V)$ such that $\lie\G (\k,\k)=0$.
Let us show that 
these null vectors remain in $\mu(\R)$ under the push-forward of $\f_s$, 
$\forall s\in\r$ (and hence the equality 
$\lie\G (\f'_s\k,\f'_s\k)=0$ holds $\forall s\in\r$ on the whole $V$). 
Recall that we assume 
that the number $m=Span\{\mu(\R)\}$ does not depend on the
point of $V$. The result is evident if $m=1$, given that then 
$\O\propto \K$ 
with $\k$ the unique direction in $\mu(\R)$ so that the hypothesis is
$\lie\K\propto\K$ which easily leads to $\f'_{s}\k\propto\k$ all over 
$V$. Thus, we will concentrate on the case with $m>1$.
Define the $m$-form $\O=\K_1\wedge\dots\wedge\K_m$ built up from 
$m$ linearly independent elements of $\mu(\R)$ 
and construct its Lorentz tensor $\S$. 
The canonical decomposition of $\R$ (see theorem \ref{DECOMP}) 
reads $\R=\beta\S+\Q$ with $\beta>0$, $\Q\in\DP^+_2(V)$, 
$\mu(\Q)\supset\mu_{\xiv}$ and $\mu(\S)=\mu(\R)$, 
so that (\ref{primera-a}) holds. Then, we know from proposition 
\ref{forms} that the property $\lie\O\propto\O$ is equivalent to equation 
(\ref{primera-b}). Since equations (\ref{primera-a}) and (\ref{primera-b}) 
lead to (\ref{partly}), we obtain via integration the following formula 
for the projector $\p=(\G+\S)/2$ over the subspace 
$Span\{\mu(\S)\}=Span\{\mu(\R)\}$ 
\be
\lie\p=(\alpha+\beta+\l)\p\hspace{5mm} \Longrightarrow\hspace{5mm}
(\f^*_s\p)\propto\p,
\label{integration}
\ee
where again equation (\ref{lie-property}) has been used. Obviously,
$\S|_x(\k,\k)=0\Longleftrightarrow \p|_x(\k,\k)=0$, whence $\forall 
\k\in\mu(\R|_x)$, $\p|_{\f_s(x)}(\f'_s\k,\f'_s\k)=\f^*_{s}\p|_{x}(\k,\k)\propto 
\p|_{x}(\k,\k)=0$, which is only possible if either $\f'_s\k$ belongs to 
$\perp Span\{\mu(\S|_{\f_s(x)})\}$ or 
$\f'_s\k\in Span\{\mu(\S|_{\f_s(x)})\}$ i.e. 
$\lie\G|_{\f_s(x)}(\f'_s\k,\f'_s\k)=0$ 
for all $s\in\r$. The former possibility is forbidden because 
the conformal invariance $\lie\O\propto\O$ implies, 
via integration by means of (\ref{lie-property}), that 
$\f'_s\k|_{\f_{s}(x)}\in Span\{\k_1|_{\f_s(x)},\dots,\k_m|_{\f_s(x)}\}=
Span\{\mu(\S|_{\f_s(x)})\}$. 
The conclusion of all this is that 
$0=\lie\G|_{\f_s(x)}(\f'_s\k,\f'_s\k)=d\f^*_s\G|_x(\k,\k)/ds$, 
$\forall\k\in\mu(\R|_x)$ and every $s\in\r$ which means that, all 
over $V$,  $\f^*_s\G(\k,\k)=0$ $\forall s\in\r$ and $\forall \k\in\mu(\R)$.
If, on the other hand, $\k$ is a null vector such that
$\lie\G|_x(\k,\k)>0$, a reasoning similar to that of the case with 
$\mu(\R)=\emptyset$ leads us to $\f^*_s\G|_x(\k,\k)>0$ if 
$\k\not\in\mu(\R|_x)$ and $s>0$, for all $x\in V$
(note that we must use the just proven property of the principal directions 
in order to be able
to show the inequality $\lie\G|_{\f_{s}(x)}(\f'_s\k,\f'_s\k)>0$ $\forall s>0$ 
if $\k\not\in\mu(\R|_x)$).

In any case, we end up proving that, for every $x\in V$,
$\f^*_s\G|_x(\k,\k)\geq 0$ for all $s>0$ 
and $\forall\k\in\partial\Z^+_{x}$, which according to theorem 
\ref{nullconvergence} is equivalent to $\f_s\in\C(V,\G)$ $\forall s>0$. 
Furthermore, we get $\mu(\R)=\mu_{\xiv}$, as desired. \N

\medskip
\noindent
{\bf Remark.} We must point out that the requirement 
$\lie\G-\alpha\G\in\DP^+_2(V)$ can be replaced by the equivalent condition 
$\lie\G(\k,\k)>0$ for every null vector field if $\mu(\R)=\emptyset$,
that is to say, the tensor $\lie\G$ fulfills 
the strict null convergence condition (see lemma \ref{convergencenull} 
for a proof of this). This last condition 
is more natural as it only involves the tensor $\lie\G$. For the case 
$\mu(\R)\neq\emptyset$ we have preferred to formulate theorem \ref{converse} 
in terms of the dominant property of the tensor $\lie\G-\alpha\G$ for a certain 
function $\alpha$, but equivalent statements using the null 
convergence condition of $\lie\G$ can also be given, see the remark after 
lemma \ref{convergencenull}.   

Theorem \ref{converse} covers all possibilities for a vector field
$\xiv$ liable to generate a one-parameter submonoid of causal 
symmetries according to the number of canonical null directions.  
This theorem together with theorem \ref{differential} comprise the necessary and
sufficient conditions which a vector field must meet in order to be
causal preserving.  Notice that we have made no hypotheses over the
differentiability properties of the metric tensor in theorem
\ref{converse} whereas the analyticity of the metric was essential to
state theorem \ref{differential}.

A proof of theorem \ref{converse} for the particular case in which $\xiv$ is 
timelike is available in \cite{LOW,H2} where some additional global properties 
of such vector fields are discussed. 
These authors require the null convergence condition for the tensor $\lie\G$ 
in order to prove that a timelike $\xiv$ is a CPVF which, 
as we have already pointed out, is equivalent to the condition 
$\lie\G-\alpha\G\in\DP^+_2(V)$ demanded in theorem \ref{converse} if 
$\lie\G(\k,\k)>0$ for every null vector $\k$. As a matter of fact, 
the proof in \cite{LOW,H2} does not depend on the timelike character 
of $\xiv$ {\em in this case}. However, for the general case with 
canonical null directions, i.e. such that the null convergence 
condition for $\lie\G$ is not strict,  
their proof can only be applied to timelike CPVFs, as they 
clearly point out, whereas the result stated 
in theorem \ref{converse} is thoroughly general and covers all possibilities. 

Theorem \ref{converse} states in particular that the canonical null 
directions of the one-parameter submonoid generated by $\xiv$ are just 
$\mu_{\xiv}=\{\k\in\d\Z^+:\ \lie\G(\k,\k)=0\}$. 
In addition, we can get the following interesting corollary.
\begin{coro}
If the vector fields $\xiv_1$ and $\xiv_2$ are causal preserving and 
$\mu_{\xiv_1}=\mu_{\xiv_2}$, then every linear combination of them 
with positive constant coefficients is causal preserving too with the 
same canonical null directions.
\label{combination}
\end{coro}
\P This is a consequence of the fact that every linear combination 
$\xiv= c_1\xiv_1+c_2\xiv_2$ with $c_1,c_{2}>0$ also satisfies the differential 
conditions of theorem \ref{differential} {\em whenever} 
$\mu_{\xiv_1}=\mu_{\xiv_2}$, and the fact that theorem \ref{converse} 
proves the sufficiency of these equations for $\xiv$.\N

\section{Conserved quantities and constants of motion}
\label{conservedquantities}
We explore next the possibility of defining conserved quantities from 
CPVFs in much the same way as it has been done with other known symmetries 
such as isometries or conformal motions.
For a general CPVF ${\bf\xiv}$ we already know (see (\ref{primera}))
that $\lie\G=\alpha\G+\U$ for some $\U\in\DP^+_2$. Then, if $\V$ 
is the vector tangent to a given curve, we get along the curve
\bnr
\nabla_{\V}(\xiv\cdot\V)&=&(\nabla_{\V}\xiv)\cdot\V+\xiv\cdot(\nabla_{\V}\V)=
\fr{1}{2}(\lie\G)(\V,\V)+\xiv\cdot(\nabla_{\V}\V)=\\
&=&\fr{1}{2}\alpha\G(\V,\V)+\fr{1}{2}\U(\V,\V)+
\xiv\cdot\nabla_{\V}\V.
\enr
 From this relation we deduce that if $\V$ is tangent to an affinely parametrized 
geodesic such that $\V|_x\in\mu(\U|_x)=\mu_{\xiv}|_{x}$ for all $x$ on the curve,
then $\xiv\cdot\V$ is constant along this null geodesic. Hence, the null 
geodesics along the canonical null directions of a CPVF 
$\xiv$ have a {\em constant} component along $\xiv$. Furthermore, 
for general affinely parametrized null geodesics it follows that 
$\xiv\cdot\V$ is monotonically non-decreasing to the future along the curve.
However, we would like to consider a class of curves more 
general than just null geodesics. The next 
definition taken from \cite{SCHOUTEN} will help us in this task.
\begin{defi}
A smooth curve $\g\subset V$ is called a {\em subgeodesic with respect to the 
vector field} $\vec{\bf p}$ if its tangent vector satisfies the equation 
$\nabla_{\V}\V=\vec{\bf p}\l+a\V$ where $\l$ is a smooth function. 
\end{defi}
We can always choose an {\em affine parametrization} for every subgeodesic
so that the term $a\V$ is removed from previous equation: $\nabla_{\V}\V=\vec{\bf p}\l$. We will assume in the sequel that we are working
with affinely parametrized subgeodesics.

Next we show how to define constants along certain of these curves.
\begin{prop}
Let $\v$ be the tangent vector of an affinely parametrized subgeodesic with 
respect to $\vec{\bf p}$. Then for a CPVF $\xiv$, the quantity 
$\xiv\cdot\v$ is constant along the subgeodesic if either: 
\begin{enumerate}
\item $\xiv$ and $\vec{\bf p}$ are orthogonal at all points of the curve and 
$\U(\v,\v)=-\alpha\v\cdot\v$.
\item $\l=-\fr{1}{2}(\xiv\cdot\vec{\bf p})^{-1}(\alpha\G+\U)(\v,\v)$ if $\xiv$ 
and $\vec{\bf p}$ are not orthogonal at some point of the subgeodesic.  
\end{enumerate}
\label{CONSERVED}
\end{prop}
\P We just replace $\nb_{\v}\v$ in the above calculation of 
$\nb_{\v}(\xiv\cdot\v)$ by its expression 
for an affinely parametrized subgeodesic, getting:
$$
\nb_{\v}(\xiv\cdot\v)=\fr{1}{2}\alpha\G(\v,\v)+
\fr{1}{2}\U(\v,\v)+\l\xiv\cdot\vec{\bf p},
$$
from what the result for each case $(i)$ and $(ii)$ follows.\N

\medskip
\noindent
{\bf Remarks.}
\begin{enumerate}
\item Conditions $(i)$ and $(ii)$ are just necessary conditions imposed upon
the CPVF $\xiv$ and the vector $\vec{\bf p}$ for $\xiv\cdot\v$ to be 
a constant.
\item Case $(i)$ of the previous proposition includes the case 
$\v\in\mu(\U)=\mu_{\xiv}$.
Observe that $\l$ is arbitrary in this case, so that for $\l=0$ we can have 
the null geodesics mentioned before.
\item The condition in $(ii)$ is very weak because {\em every} vector field 
$\vec{\bf p}$ non-orthogonal to $\xiv$ gives rise to a constant of motion 
along the affine subgeodesics with respect to $\vec{\bf p}$ by simply 
choosing an adequate $\lambda$ proportional to $(\alpha\G+\U)(\v,\v)$
(this is similar to the case studied in \cite{KERR-SCHILD}).
\end{enumerate}

The construction of conserved currents is also possible for CPVFs as we 
prove in the next proposition.
\begin{prop}
Let $T_{a_1\dots a_p}$ be a totally symmetric and traceless rank-$p$ tensor and 
assume further that $\nb_bT^b_{\ a_2\dots a_p}=0$. Then for a CPVF 
$\xiv$ the current $j^b=T^b_{\ a_2\dots a_p}\xi^{a_2}\dots \xi^{a_p}$ is 
divergence-free if $T_{bc a_{3}\dots a_p}U^{bc}=0$.
\label{current}
\end{prop} 
\P A direct calculation using the assumptions on $T_{a_1\dots a_p}$ gives
$$
\nb_bj^b=(p-1) T^{bc}{}_{\ a_3\dots a_p}\nb_{(b}\xi_{c)}\xi^{a_{3}}
\ldots\xi^{a_{p}}=\frac{p-1}{2}\, T^{bc}{}_{\ a_3\dots a_p}U_{bc}
\xi^{a_{3}}\ldots\xi^{a_{p}}
$$ 
where (\ref{primera}) has been used for $2\nb_{(b}\xi_{c)}=\lie\G_{bc}$.
The result is now trivial.\N

Observe that the condition of the proposition is sufficient, but it 
will be enough with just $T^{bc}{}_{\ a_3\dots a_p}U_{bc}
\xi^{a_{3}}\ldots\xi^{a_{p}}=0$.

Relevant examples of tensors with the properties stated in proposition 
\ref{current} spring to mind at once as for instance every traceless Einstein 
tensor, the energy-momentum tensor of a source-free electromagnetic field in four 
dimensions, perfect fluids whose equation of state is $p=\rho/(n-1)$,
and many superenergy tensors \cite{SUP} like the Bel-Robinson tensor in 
Einstein spaces \cite{BEL} or the trace of the Chevreton tensor for 
electromagnetic fields \cite{CHEVRETON}. In any case one must check the 
condition $T_{ab\dots a_p}U^{ab}=0$ in order to actually obtain a conserved 
current which in general is not a trivial matter and requires a case-by-case 
discussion. We present next some examples of physical 
relevance. 

\paragraph{The electromagnetic field}
\  

 The simplest tensor one can think of complying with the above conditions is the 
energy-momentum tensor of any source-free electromagnetic field in four 
dimensions. This is a traceless rank-two tensor $T_{ab}$ which
is divergence-free if Maxwell equations are satisfied and hence 
$j^a=T^a_{\ b}\xi^b$ will be conserved for a CPVF $\xiv$
if $T_{ab}U^{ab}=0$. 

When the electromagnetic field is null
then $T_{ab}=k_ak_b$ where $\K$ is the single principal null direction of the 
Maxwell field. Therefore, $T_{ab}U^{ab}$ will vanish if $\k\in\mu_{\xiv}$ or,
in other words, if $\k$ is a canonical null direction of the submonoid of causal 
symmetries generated by $\xiv$ (note that in this case 
$T_{ab}$ is traceless for any dimension $n$). 

Another possibility arises when $U_{ab}$ is
proportional to a Lorentz tensor $S_{ab}$ commuting with $T_{ab}$, 
and the electromagnetic field is non-null. 
The square of the (symmetric) tensor $T_{ap}U^p_{\ b}$, in the sense of 
proposition \ref{cross-mu}, is proportional to the metric tensor ---due 
to the Rainich conditions for $T_{ab}$ in four dimensions, see e.g. 
PI---, whence theorem 6.1 of PI implies that $T_{ap}U^p_{\ b}$ must be
proportional to the superenergy tensor of a simple form. As its trace 
vanishes owing to the condition $T_{ab}U^{ab}=0$, the generalized Rainich 
conditions of PI tell us that the simple form must be a 2-form from what we 
deduce, working in the common orthonormal basis of both $T_{ab}$ and 
$U_{ab}$, that $U_{ab}$ is also the superenergy tensor of a simple 2-form,
different from the ones generating $T_{ab}$ and $T_{ap}U^p_{\ b}$. 
Writing $U_{ab}=\beta S_{ab}$ it is easy to prove by Lie derivating 
the relation $\rmg_{ab}=S_{ap}S^p_{\ b}$ that the vector $\xiv$ 
is a bi-conformal vector field (see just before proposition \ref{n-1}). 
Summarizing, the current $j^b=T^b{}_{c}\xi^c$ will be conserved if 
$\xiv$ is a bi-conformal vector field with $\mu_{\xiv}\neq \mu(\T)$ but such 
that $Span\{\mu_{\xiv}\}\cap Span\{\mu(\T)\}$ defines a common 
timelike eigendirection for $\T$ and $\U$.

\paragraph{The Bel-Robinson tensor}
\ 

 Another example of tensor satisfying the main hypotheses of proposition 
\ref{current} is the Bel-Robinson tensor in an Einstein space $(n=4)$. 
This tensor was introduced independently by Bel \cite{BEL} and 
Robinson \cite{ROBINSON} in General Relativity as a gravitational analog of 
the energy-momentum tensor of the electromagnetic field and its expression 
reads
$$
{\cal T}_{abcd}=C_{apbq}C_{c\ d\ }^{\ p\ q}+(*C)_{apbq}(*C)_{c\ d\ }^{\ p\ q},
$$
where $*C$ stands for the (unique) dual of the Weyl tensor $C_{abcd}$. The 
Bel-Robinson tensor is a completely symmetric, traceless and 
divergence-free {\em future} tensor \cite{RINDLER,SUP}. Generalizations of this 
construction to the so-called superenergy tensors are presented in \cite{SUP,PP}
(actually one can build from these objects more tensors satisfying the 
requirements of proposition \ref{current}). The remaining condition 
${\cal T}_{abcd}U^{ab}=0$ of proposition \ref{current} can be achieved 
depending on how the future tensor $\U$, or equivalently 
$\mu_{\xiv}$, is related to the set $\mu({\cal T})$ which, as
is well known, consists of the principal null
vectors of the Weyl tensor (see e.g.\cite{RINDLER,GORAN}). Several 
possibilities leading to a conservation law are:
\begin{itemize}
\item
If $U_{ab}=k_{(a}u_{b)}$ with $k_a$ null and $u_a$ timelike, so that  
$\mu(\U)=\mu_{\xiv}=\{\l\k\}$, $\l>0$ if $\k$ is future pointing and 
$\xiv$ is a degenerate CPVF. In this case the condition becomes
$$
{\cal T}_{abcd}U^{ab}=0\Rightarrow {\cal T}_{abcd}k^au^b=0,
$$
which entails both $0={\cal T}_{abcd}k^a$ and the spacetime being of Petrov 
type N (theorem 2 of \cite{GORAN}) with $\k$ as 
the unique repeated principal null direction of the Weyl tensor. This 
holds true, in fact, for all causal CPVFs if the set of its 
canonical null directions includes the principal null direction
of the Petrov type N Weyl tensor, because then ${\cal T}_{abcd}\propto 
k_{a}k_{b}k_{c}k_{d}$ and obviously ${\cal T}_{abcd}U^{ab}=0$. Thus, in 
any Petrov type N spacetime with $\k$ as the repeated principal null direction,
all causal CPVFs $\xiv$  such that $\mu_{\xiv}\ni \k$ provide 
conserved currents ${\cal T}^a{}_{bcd}\xi^b\xi^c\xi^d$.
\item Another possibility is that $U_{ab}=k_ak_b$ with $\k$ null, so 
that $\xiv$ is a degenerate CPVF. In this case the condition
becomes ${\cal T}_{abcd}k^ak^b=0$, which again implies that $\k$ is a repeated 
null direction of the Weyl tensor but now the spacetime can also be of
Petrov type III (apart from type N), see \cite{GORAN}.
\end{itemize}

\medskip
\noindent
{\bf Example 3.} In this example we present a spacetime $(V,\G)$
where a nontrivial conserved current can 
be explicitly constructed in the fashion described above. The line element is:
$$
ds^2=\fr{2}{x^2}\!\left[(az^2+b)dt^2-\fr{dz^2}{az^2+b}-\fr{dx^2}{x^2(-a+lx-2e^2x^2)}-
x^2(-a+lx-2e^2x^2)dy^2\right]
$$
where $a,b,e,l$ are constants and $a>0$.
This is a type-D Einstein-Maxwell solution \cite{PLEBANSKI,KRAMERS} whose 
energy-momentum tensor takes the form ($8\pi G=c=1$)
$$
T^a{}_{b}=e^2 x^4 \, \mbox{diag}(1,1,-1-1)
$$
so that $e$ can be interpreted as the charge creating the electromagnetic 
field.

The one-parameter group of diffeomorphisms generated by $\xiv=\d/\d z$ 
contains a submonoid of proper causal symmetries
in the half space $z\geq 0$ because, for $s>0$:
$$
\DP^+_{2}(V) \ni \f^*_s\G=\left[\fr{a(z+s)^2+b}{az^2+b}-1\right](\z^0)^2+
\left[1-\fr{az^2+b}{a(z+s)^2+b}\right](\z^1)^2+\G ,
$$
where we have used the orthonormal cobasis dual to the set of eigenvectors of 
$T^a_{\ b}$.
Therefore we have for the Lie derivative of $\G$
$$
\lie\G=\U=\left(\fr{2az}{az^2+b}\right)\left[(\z^0)^2+(\z^1)^2\right]
$$
so that this is an {\em aligned} CPVF with 
$\mu_{\xiv}=\emptyset$.
 We can check that $U^{ab}T_{ab}=0$ so that the vector field
$j^a=T^a_{\ b}\xi^b$ is divergence-free. The explicit computation of such 
current gives $\vec{\bf{j}}=e^2x^4\d/\d z$.

\section{Conclusions}
We have found the necessary and sufficient conditions for a vector 
field to be causal preserving, that is to say, to locally generate a 
one-parameter submonoid ---it cannot be a group unless in the known 
case of conformal Killing vectors---of causal transformations. This implies that 
the causal future-directed vector fields are preserved along the flow 
of the causal preserving vector field. These transformations and its 
infinitesimal generators generalize straightforwardly the corresponding conformal 
counterparts. The generalization arises because the new vector fields 
preserve the whole solid Lorentz cone but {\em only part} of the null cone 
remains on the null cone. This part is the ``conformal'' part of the 
transformation, so to speak. Thus, a classification of causal 
preserving vector fields can be immediately performed, going from the 
conformal Killing vectors (the full null cone is preserved), passing 
through the case where only one null direction is not preserved 
---these are a particular case of bi-conformal vector fields, see 
\cite{PIV}---, and so on until one reaches the so-called degenerate case 
for which only one null direction is preserved ---including the 
previously studied Kerr-Schild symmetries \cite{KERR-SCHILD}--- and ending with 
the case where no null directions are kept; in this last case, there 
may still be one timelike direction which is maintained invariant 
---the aligned case.

Causal preserving vector fields seem to have relevant applications. 
In the paper we have presented a few. To start with, they provide 
constants of motion along preferred null geodesics, or more generally, 
along subgeodesics. And if not, they give monotonically increasing 
quantities along the mentioned curves. The importance of subgeodesics in physical 
applications may arise if a canonical, preferred, or otherwise 
distinguished, global vector field is defined. For 
instance, one can choose a particular observer defined by a timelike 
vector field $\u$. Then, obviously, subgeodesics relative to $\u$ are 
physically meaningful as they are can be seen as autoparallel curves
of a (non-metric) connection associated to this 
observer. As a matter of fact, the so-called chorodesics, see e.g. 
\cite{BELLOSA,BELLOSA2} and references therein, are just a particular 
case of subgeodesics. For other applications of subgeodesics the 
reader is referred to \cite{SCHOUTEN}. The physical or geometric 
interpretation of these constants of motion is similar to that 
arising for Killing vectors: the component along the CPVF of the 
vector tangent to the (sub)geodesic is constant along the curve. More 
generally, if these components are not constant, in many cases one 
can derive monotonicity properties for them, a result which may be 
relevant in applications of causality theory or global geometry.

Furthermore, the CPVF can be used to construct divergence-free vector 
fields, currents in short, in a manner analogous to that of conformal 
Killing or Killing vectors. The importance of these currents, of 
course, is that they provide conserved quantities for the field 
involved which are in principle independent of those previously known. 
We have presented several particular cases for the 
electromagnetic field or the Bel-Robinson gravitational 
``superenergy''. The true relevance of these quantities is still to be 
investigated.

Finally, from a pure geometrical viewpoint, the causal preserving 
vector fields can be used to characterize splitting of the Lorentzian
manifolds---such as in the case of warped product spacetimes and the 
like--- or other generalizations of conformal relations
\cite{LETTER,H2,LOW,PIV}, as well as to analyze further the causal 
structure and isocausality of Lorentzian manifolds in the sense of 
\cite{PII} by using infinitesimal techniques. This is under current 
study.

\section*{Acknowledgements}
This work has been carried out under financial support of the Basque Country 
University and the Spanish Ministry of Science and Technology under the projects 
9/UPV00172.310-14456/2002 and BFM 2000-0018, respectively.

\appendix

\section*{Appendix A: Essentials of causal relationship and causal tensors}
\setcounter{section}{1}
\setcounter{equation}{0}
\def\thesection{\Alph{section}}
We review next the basic tools presented in PII and PI 
(refs.\cite{PII,PI}) which are needed in this paper. 
 A vector $\v$ in the tangent space of each point of a Lorentzian manifold is 
classified as spacelike or causal according to whether the sign of the scalar 
product $\G(\v,\v)$ is negative or not
 and causal vectors are further divided into null ($\G(\v,\v)=0$) and timelike 
($\G(\v,\v)>0$) ones. If we choose a causal vector $\v_1\in T_{x}(V)$ to
be future pointing, the 
remaining causal vectors are termed as future directed with respect to $\v_1$ if 
$\G(\v,\v_1)>0$ and past directed otherwise. Of course this set remains the
same if we replace $\v_1$ by another future directed vector $\v_2$ with
respect to $\v_1$ making thus possible to define the set of causal
future directed vectors which, following PII, will be denoted by $\Z^+_x$. This
set has also a past counterpart $\Z^-_x$ (we will omit the past duals, denoted in 
general with a ``-'', when we define and use other causal objects in this paper).
The boundary $\d\Z^+_x$ of $\Z^+_x$ is the set of null future directed vectors. 

These definitions extend
straightforwardly to the bundle $T(V)$ and so we can speak of spacelike and 
future-directed causal vectors over $V$ with the notation $\Z^+(V)$ for 
the latter set. The set $\Z(V)$ is the union of $\Z^+(V)$ and $\Z^-(V)$.
 A Lorentzian
manifold is causally orientable if it admits an everywhere
 continuous future-directed vector field (i.e. a section of $\Z^+(V)$). 
We will assume that all
our Lorentzian manifolds are causally orientable. We will implicitly assume that 
the causal vectors are future-directed unless otherwise stated.

The previous construction can be repeated for 1-forms
if we work with the contravariant form $\rmg^{ab}$ of the metric tensor. 
Furthermore we can even extend it to higher rank tensors by means of the following 
definition \cite{SUP,PI}
\begin{defi}
A tensor $\T\in T^{0}_r(x)$ is said to be future (respectively past) if 
$\T(\u_1,\dots,\u_r)\geq 0$ (resp.$\leq 0$) $\forall$ $\u_1,\dots,\u_r\in\Z^+_x$.
Future and past tensors are called causal tensors. 
\label{CAUS-TENSR}
\end{defi}
The set of future tensors at $x$ will be denoted by $\DP^+_r|_x$
with the obvious notation for causal tensors over the whole manifold. We will
sometimes omit the set over which we are considering causal tensors in order
to alleviate the notation. The symbol
$(\DP^+)^s_r$ will be used for future tensors defined on $T^s_r(V)$ and 
$\DP^+$ stands for the set of all future tensors. As was
proved in PI, definition \ref{CAUS-TENSR} can be equivalently stated by
(i) just demanding that $\T(\k_1,\dots,\k_r)\geq 0$ for all future-directed
{\em null} vectors $\k_1,\dots,\k_r$ or (ii) requiring that 
$\T(\u_1,\dots,\u_r)> 0$ for all future-directed {\em timelike} vectors 
$\u_1,\dots,\u_r$. Other interesting criteria to ascertain when a 
given tensor is in $\DP^+_r$ are given in \cite{SUP,PI,PII},
as well as some interesting properties of these tensors. 
We can think of a causal tensor as a higher rank generalization of the
tensors satisfying the well known dominant energy condition used in 
General Relativity (future tensors are also called {\it dominant tensors}). 

We present next a definition (PII) which will play a very important role along 
this work.
\begin{defi}
The {\em principal directions} of $\T\in\DP^+_r|_x$
are the vectors $\u\in\Z^+_x$ such that $\T(\u,\dots,\u)=0$.  
The set of principal directions of $\T$ is denoted by $\mu(\T)$.
\label{CAN-DIR}
\end{defi}
Of course we can define the principal directions for arbitrary elements of 
$\DP^{+}_r(V)$ (or its sections) which will consist of elements of 
$\Z^+(V)$. Clearly, for an element $\T\in\DP^+_r(V)$, the set of its principal 
directions may change from point to point and if $\T$ is a continuous 
(differentiable) section of $\DP^+_r(V)$
we will be able to construct continuous (differentiable) vector fields $\v$
such that $\v|_{x}\in\mu(\T|_x)$, $\forall x\in V$. As we are going 
to show, these principal directions have a great deal to do with the algebraic 
classification of the tensor under study.

To start with, the principal directions are necessarily null
($i_{\k}$ stands for the usual inner contraction with $\k$).
\begin{prop}
 A vector $\k\in\Z^+_{x}$ is in $\mu(\T)$ for some $\T\in\DP^+_r|_x$, ($r>1$) 
iff it is null and belongs to $\mu(i_{\k}\T)$. 
In the case of $r=1$ the condition is $\K\propto\T$.   
\label{CONTRACTION}
\end{prop}
\P Clearly $\k\in\mu(\T)\Leftrightarrow\k\in\mu(i_{\k}\T)$. On the other hand
 if $\k$ is in $\mu(\T)$ then the future-directed 1-form 
$v_{a}\equiv T_{ab_2\dots b_r}k^{b_2}\dots k^{b_r}$ is orthogonal to $k^a$ 
which is only possible if $k^a$ and $v_a$ are null and proportional 
(see property 2.2 of PI for an explanation about the causal character of $v_a$).
When $r=1$ then $\k\in\mu(\T)\Leftrightarrow k^aT_a=0$ which again only happens if
both $\T$ and $\K$ are null and proportional.\N

The set $\mu(\T)$ is never a vector space. If this set consists of a single 
linearly independent element then it is clearly a one-dimensional semi-space 
formed by the positive multiples of that element. 
In the case of having two linearly independent elements, say $\k_1$ and $\k_2$, 
we deduce that $\mu(\T)=\{\l_1\k_1,\l_2\k_2\}$, $\l_1,\l_2\in\r^+$ since the 
linear combination of two null vectors can never be null.

The idea of principal direction has been already used in the literature
mainly to classify algebraically the electromagnetic and Weyl tensors in 
Relativity. It is well known \cite{RINDLER,GORAN} that these principal
directions are (in four dimensions) precisely the principal
directions according to definition \ref{CAN-DIR} of the energy-momentum tensor 
of the electromagnetic field and of the Bel-Robinson tensor, 
respectively, being both these tensors elements of $\DP^+$.
See \cite{PP} for a recent study of the principal directions of more general
elements of $\DP^+$. If $\T\in\DP^+_2|_x$ the calculation of $\mu(\T)$ is 
accomplished by means of the next lemma proven in PII:
\begin{lem}
For any $\T\in\DP^+_2|_x$, $\k\in \mu(\T)$ iff $\k$ is a null 
eigenvector of $\T$.\N
\label{N-N}
\end{lem}  
As a corollary of this we can give a result for higher rank tensors.
\begin{coro}
For $\T\in\DP^+_r|_{x}$, $r\geq 4$, $\k\in\mu(\T)\Leftrightarrow$ $\k$ 
is a null eigenvector of $\overbrace{i_{\k}\dots i_{\k}}^{r-2}\T$.
\end{coro}
\P Just apply recursively proposition \ref{CONTRACTION} $r-2$ times
to end up with a rank-2 tensor and then use lemma \ref{N-N}.\N

 For a rank-2 causal tensor the number of its linearly independent principal
directions is the dimension
of the subspace $Span\{\mu(\T)\}$ which is always an eigenspace of the tensor 
$\T$ for one of its positive eigenvalues if $\T$ is symmetric (otherwise
 the scalar product of {\it null} vectors corresponding to 
different eigenvalues would have to be zero which is impossible). 
To get a better picture of this we show next a table with the allowed 
Segre types for symmetric rank-2 causal tensors together with 
the principal directions of each type.
\begin{table}[h]
\vspace{.4cm}
\begin{tabular}{|l|ll|}
\hline\hline\hline\hline
$dim(Span\{\mu(\T)\})$ & \multicolumn{2}{c|}{\em Segre type}\\ \hline
0  & $[1,1\dots 1]$ & and its spatial degeneracies \\
1 (degenerate type) & $[2\ 1\dots 1]$ & and its degeneracies\\
2 & $[(1,1)1\dots 1]$ & and its spatial degeneracies \\
.........& ..................&.....................................\\
$n-1$ & $[(1,1 1\dots 1) 1]$ &\ \ \\ 
$n$   & $[(1,1 1\dots 1)]$   &\ \                        \\ \hline\hline
\end{tabular}  
\caption{Possible Segre types for a rank-2 symmetric causal tensor. It is 
understood that the spatial degeneracies do not coincide with the timelike ones.}
\label{Segre}
\end{table}

\noindent
A justification of this result for four dimensions can be 
found in many textbooks (see e. g. \cite{KRAMERS,HAWKING}) although its 
generalization
for $n$ dimensions is quite straightforward (see e.g. theorem 1 of \cite{HOGLUND})
and it shall not be presented here.

The eigenvalue associated to $Span\{\mu(\T)\}$ has multiplicity 
dim$Span\{\mu(\T)\}$ (this is the degenerate eigenvalue with round brackets 
in the Segre notation of table \ref{Segre}). The so-called degenerate type 
possesses a single null eigenvector and is the only type with no orthonormal 
basis formed by eigenvectors. Each rank-2 symmetric causal tensor has 
a canonical form according to its Segre type being these forms further 
constrained once the dominant property is imposed.
\begin{prop}
A rank-2 symmetric tensor $\T$ is an element of $\DP^+_2|_{x}$ if and 
only if its 
algebraic type is one of table \ref{Segre} and its eigenvalues satisfy
the following properties:
\begin{itemize}
\item the eigenvalue $\l_0$ associated to the timelike part is 
greater than or equal to the absolute value of the remaining eigenvalues (spacelike
eigenvalues). 
\item If $\T$ is of Segre type $[2\ 1\dots 1]$ or its degeneracies then it
 can be written as 
$\T=\T_0+\l\K\otimes\K$ with
$\T_0\in\DP^+_2$ a nondegenerate type other than $[1,1\dots 1]$ and $\K$ the 
single null eigenvector of $\T$ (which is a null eigenvector of $\T_0$ as well). 
The conditions are now those necessary for $\T_0$ plus $\l>0$.
\end{itemize} 
\label{DP-condition}
\end{prop}  
\P The possible algebraic types have been already discussed and the first 
statement of this proposition is a particular case of a more general result 
proven in lemma 4.1 of \cite{SUP}. The second statement is needed in the proof 
of theorem 4.1 of PI and its proof is supplied there (see also theorem 2 of 
\cite{HOGLUND}).\N 

The set of null eigenvectors can also be used to write the rank-2 symmetric
causal tensors in a certain canonical form which is used in this work
and was proven in PI.
\begin{theo}
Every symmetric ${\bf T}\in\DP^{+}_2|_x$ can be written canonically as the sum
${\bf T}=\sum^{n}_{r=1}{\bf T}\{{\bf \O}_{[r]}\}$ of rank-2
``super-energy tensors'' ${\bf T}\{{\bf \O}_{[r]}\}\in \DP^+_2|_x$ of
simple $r$-forms ${\bf \O}_{[r]}$. Furthermore, the decomposition is
characterized by the null eigenvectors of ${\bf T}$ as follows:
if ${\bf T}$ has $m$ linearly
independent null eigenvectors $\k_1, \dots, \k_m$ then the sum starts
at
$r=m$ with ${\bf \O}_{[m]}={\bf k}_1\wedge\dots\wedge {\bf k}_m$ and the remaining
 simple forms $\O_{[r]}$ contain $\O_{[m]}$ as a factor; if
${\bf T}$
has no null eigenvector then the sum starts at $r=1$ with
$\vec{\O}_{[1]}$ 
the timelike eigenvector of ${\bf T}$ and again $\O_{[1]}$ is a factor 
of all the simple forms $\O_{[r]}$.\label{CLASS}\N
\label{DECOMP}
\end{theo}
Let us recall that the superenergy tensor of a $r$-form $\Sg$ is given 
by\footnote{Clearly $\T\{\Sg\}$ is a generalization to arbitrary degree forms
of the energy-momentum tensor of the electromagnetic field in terms of the 
Faraday 2-form ${\bf F}$. This generalizing idea can
be pursued further and one can define superenergy tensors of more general objects
than antisymmetric tensors \cite{SUP,PP}.}:
\be
T_{ab}\{\Sg\}=\fr{(-1)^{r-1}}{(r-1)!}
\left[\Sigma_{aa_2\dots a_r}\Sigma_{b}^{\ a_2\dots a_r}-
\fr{1}{2r}\rmg_{ab}\Sigma_{a_1\dots a_n}\Sigma^{a_1\dots a_n}\right].
\label{s-e-f}
\ee
If the form $\Sg$ is simple, i. e., it is the wedge product of $r$ 1-forms
${\bf u}_1,\dots,{\bf u}_r$,  
then the tensor $T^{a}_{\ b}$ is proportional to an involutory Lorentz 
transformation hence $T_{ab}T^{b}_{\ c}=\rho\rmg_{ac}$ and
all the vectors $\{\u_k\}_{k=1,\dots,r}$ are
eigenvectors of $T^{a}_{\ b}$ with the same eigenvalue (PI). Indeed,
the tensor $T^a_{\ b}$ has only one or two different eigenvalues depending on
whether $\rho$ vanishes or not. In the non-vanishing case $T^a_{\ b}$ 
can be brought into the diagonal form with 
$Span\{\u_1,\dots,\u_r\}=Span\{\mu(\T)\}$ and 
its orthogonal space as the only eigenspaces of the two different eigenvalues 
whereas for $\rho$ vanishing $T^{a}_{\ b}$ can never be diagonalized and all
the eigenvectors have 0 as eigenvalue (see next paragraph). Every null vector 
entering as a factor in a simple $\Sg$ is an element of $\mu(\T\{\Sg\})$ so 
in the previous theorem $\mu(\T)$ is always a subset of the set of principal 
directions of each term of the decomposition. An account of this and other 
properties of these tensors can be found in PI.

The simple $r$-form $\Sg$ can be normalized with 
$\Sigma_{a_1\dots a_r}\Sigma^{a_1\dots a_r}=(-1)^{r-1}2r!$ so as to get an 
involutory Lorentz transformation $S^a_{\ b}=T^a_{\ b}\{\Sg\}$ as its 
superenergy tensor provided that $\Sg$ is not null. If $\Sg$ is null, 
it takes the form $\Sg=\K\wedge{\bf v}_1\dots\wedge{\bf v}_{r-1}$ where 
${\bf v}_1,\dots,{\bf v}_{r-1}$ are spacelike vectors and $\K$ is null 
(this is the case with $\rho=0$), so that $\T\{\Sg\}=f\K\otimes\K$.
Involutory Lorentz transformations shall be referred to in this work as 
{\em Lorentz tensors} and they are always the superenergy tensor of a certain 
simple form as was explicitly proven in PI (theorem 6.1) so there exists a 
surjective map between Lorentz tensors and 
appropriate normalized simple forms. This relation is not injective though,
as the simple normalized forms $\pm\O$ and $\pm^*\O$ give rise to the same 
Lorentz tensor $\T\{\O\}$, where the star stands for the Hodge dual. 
Therefore, we deduce that each Lorentz tensor has, up to duality and sign, 
an associated simple form given by (\ref{s-e-f}). 

If $\Sg$ is a simple $r$-form as the one
defined above, then $^*\Sg$ is also simple (a form is simple iff its dual is
also simple),  and formed by the wedge product of $n-r$ 1-forms which are a basis
of $\perp Span\{{\bf u}_1,\dots,{\bf u}_r\}$ if $\Sg\cdot\Sg\neq 0$ or 
$\perp Span\{\K,{\bf v}_1,\dots,{\bf v}_{r-1}\}$ for the $\rho=0$ case. 
Hence a superenergy rank-2 tensor can always be viewed as the superenergy of 
a $r$-form or the superenergy of the dual $(n-r)$-form and every property stated
in terms of the $r$-form admits a dual description stemming from
the dual $(n-r)$-form.
                            
If $\rho\neq 0$ we can get a Lorentz tensor 
$S^a_{\ b}$ as already  explained. From this we can define
the tensor $P_{ab}=\fr{1}{2}(\rmg_{ab}+S_{ab})$ which will be useful 
in the main text. Straightforward  properties of 
this tensor are $P_{ac}P^c_{\ b}=P_{ab}=P_{(ab)}$, $P^a_{\ a}=r$
and det($P_{ab}$)=0 if $S_{ab}\neq\rmg_{ab}$ from what we conclude 
that $P^a_{\ b}$ is an orthogonal projector. The next result 
identifies the subspace to which it projects.
\begin{prop}
$P^a_{\ b}$ is an orthogonal projector onto 
$Span\{\mu(\S)\}$.
\label{projection}
\end{prop}
\P As we explained before, $Span\{\mu(\S)\}$ and  
$\perp\!\!\!\!Span\{\mu(\S)\}$ are the two eigenspaces associated to the only two 
eigenvalues of $S^a_{\ b}$ which in this case are +1 and -1 respectively 
as it is obvious from the property $S_{ac}S^c_{\ b}=\rmg_{ab}$.
 Since $S^a_{\ b}$ can be diagonalized,
 the direct sum of the
above two spaces must be equal to the entire vector space which means that 
 by construction $P^{a}_{\ b}$  has 
$Span\{\mu(\S)\}$ as the eigenspace with 
+1 eigenvalue.\N    

 From the proof of this proposition, we deduce that either $Span\{\mu(\S)\}$ 
or its orthogonal complement must be timelike. The fact that 
$S_{ab}\in\DP^+_2(V)$ implies that $S^a_{\ b}$ maps future-directed vectors 
onto future-directed vectors so the timelike space is associated to the positive 
eigenvalue. Moreover, in view of the previous theorem we deduce that 
dim$(Span\{\mu(\T)\})=n \Leftrightarrow\T\propto\G$
(this can also be seen from table \ref{Segre}). Therefore every rank-2 causal 
tensor with $n$ linearly independent principal directions is up to a factor 
the metric tensor.

It is possible to enlarge the set of principal directions of a rank-2 causal
tensor $T_{ab}$ if we search for the set of vectors in $\d\Z^+|_x$ which are 
mapped to null vectors under the endomorphism $T^a_{\ b}$.
\begin{defi}
The set of null directions of $\T\in\DP^+_2|_x$ is defined as follows:
\[
\sigma(\T)=\{\k\in\d\Z^+_x:T^a_{\ b}k^b\in\d\Z^+_x\}
\]
\label{SIGMA}
\end{defi}
Clearly every pair $k^a$, $l^a$ of future-pointing 
null vectors such that $T_{ab}k^al^b=0$ comply 
with the previous definition. 
Furthermore, if $l^a=T^a_{\ b}k^b$ for $k^b\in\s(\T)$ then
$l_aT^a_{\ b}=\l k_b$, $\l>0$ as follows from the property 
$0=l_al^a=T_{ab}l^ak^b$. Every principal direction of $\T$ is also a null 
direction although the converse is not true in general. Actually $\s(\T)$ 
can be worked out in terms of the principal directions of another tensor 
related to $\T$ (recall from PI that if $M_{ab}$ and $N_{ab}$ are 
tensors then their $1-1$ inner product is denoted as
$(M\ _1\times_1N)_{ab}=M_{ca}N^{c}_{\ b}$. However, most of our applications 
will deal with symmetric causal tensors in which case we 
can drop the subindexes in the notation $\ _1\times_{1}=\times$, as in PI.)
\begin{prop}
For every $\T\in\DP^+_2(V)$ we have that $\s(\T)=\mu(\T\ _1\times_1\T)$
\label{cross-mu}
\end{prop} 
\P Pick up any $\k\in\s(\T)$ and define the future directed null vector
 $n^a\equiv T^{a}_{\ b}k^b$. Then $0=n_an^a=T_{ab}k^bT^a_{\ c}k^c=
(T\ _1\times_1 T )_{bc}k^bk^c\Rightarrow\k\in\mu(\T\ _1\times_1\T)$. Conversely
if $0=(T\ _1\times_1 T)_{ab}k^ak^b$ then, given that $\T\in\DP^+_2|_x$ we get 
that necessarily $T^a_{\ b}k^b$ must be future directed and null, 
proving thus that $\k\in\s(\T)$.\N

Another case of special relevance occurs when the set
of principal directions is the same as the set of null directions. For 
instance we have
\begin{prop}
Let $\T$ of $\DP^+_2|_x$ be symmetric. Then
\begin{enumerate}
  \item If $\T$ is degenerate, then $\s(\T)=\mu(\T)$ iff $T_{0}\neq 0$, 
with $T_{0}$ as in proposition \ref{DP-condition}.
  \item If $\T$ has Lorentzian signature then $\s(\T)=\mu(\T)$.
\end{enumerate}
\label{lorentzian}
\end{prop}
\P For (i), we can write $\T=\T_0+\l\K\otimes\K$, $\l>0$, for a null $\K$ with 
$\mu(\T)=\{\rho\k\}$, $\rho>0$. Therefore, for every $\n\in\s(\T)$ we have 
that the linear combination of future-directed vectors 
$T^a_{0\ b}n^b+\l(k_bn^b)k^a$ must be null and future-directed,
which implies that $T^a_{0\ b}n^b\propto k^a$
or in other words $0=T_{0\ ab}k^an^b$ and thus, if $\T_{0}\neq 0$,
$T_{0\ ab}k^b\propto n_a$. But we know that $\k$ 
is a null eigenvector of $\T_0$ with positive eigenvalue (see again 
proposition \ref{DP-condition}), hence $\k\propto\n$ and we conclude that 
in this case $\s(\T)=\mu(\T)=\{\rho\k\}$, $\rho>0$. On the other hand, 
if $\T_{0}=0$ then $\sigma(\T)=\partial\Z^+_{x}\neq \mu(\T)$ so the 
result follows. To prove (ii), observe first that if $\T$ is 
degenerate and has Lorentzian signature, then necessarily 
$\T_{0}\neq 0$ so this case is covered by (i) and we can concentrate 
only on non-degenerate $\T$. In this case, $T^a_{\ b}$ has an
orthonormal basis of eigenvectors and its signature is Lorentzian if and only if 
all the eigenvalues are positive. Thus, if we write $\T$ in that 
orthonormal basis and compute $\mu(\T\times\T)$ we get at once that 
$Span\{\mu(\T)\}=Span\{\mu(\T\times\T)\}$ which implies $\mu(\T)=\mu(\T\times\T)$
and thus $\mu(\T)=\s(\T)$. \N 

Another result which involves the set of null directions and the set of 
principal directions is the following one.
\begin{lem}
Let $\T_1$ and $\T_2$ be symmetric nondegenerate elements of $\DP^+_2|_x$ with
the property $\s(\T_1)=\s(\T_2)=\mu(\T_1)=\mu(\T_2)$. Then we can find a 
positive constant $\alpha$ such that $\T_1-\alpha\T_2\in\DP^+_2$.
\label{basic-lemma}
\end{lem}
\P We divide up in cases depending on the number of linearly independent 
principal directions. Suppose first that $2\leq m$=dim$(Span(\mu(\T_1)))= 
$dim($Span(\mu(\T_2))$).
 Then  $T_x(V)=Span(\mu(\T_1))\oplus(Span(\mu(\T_1)))^\perp$ where due to the
hypotheses, $Span(\mu(\T_1))^\perp$ is a $(n-m)$-dimensional spacelike subspace of $T_x(V)$
which is invariant under the endomorphisms $T^{a}_{\!1\ b}$ and $T^{a}_{\!2\ b}$
so we can find an orthonormal basis in which $\T_1$ and $\T_2$ take the form
\be
\T_1=\left(\begin{array}{cc}
               {\mathbb I}_{\l_1}& \\
                                  & {\bf A_1}      
\end{array}\right),\ 
                  \T_2=          \left(\begin{array}{cc}
                                     {\mathbb I}_{\l_2}& \\
                                                        & {\bf A_2}
\end{array}\right).
\label{canonical}
\ee             
${\mathbb I}_{\l_1}$ and ${\mathbb I}_{\l_2}$ are $m\times m$ square matrices 
 given by ${\mathbb I}_{\l_1}$=diag($\l_1,-\l_1,\dots,-\l_1$), 
${\mathbb I}_{\l_2}=$diag($\l_2,-\l_2,\dots,-\l_2$) whereas ${\bf A_1}$ and 
${\bf A_2}$ are {\it symmetric} bilinear forms acting on a vector space equipped
with an Euclidean scalar product. This means that we can find an orthonormal basis 
 with respect to this scalar product which diagonalizes both ${\bf A}_1$ and 
${\bf A}_2$ from what we conclude that there exists an orthonormal basis for 
the Lorentzian scalar product bringing both $\T_1$ and $\T_2$ into their
canonical diagonal form. Thus in this new basis we get the above written expressions for
$\T_1$ and $\T_2$ where now ${\bf A}_1$=diag($a_1,\dots,a_{n-m}$) and 
${\bf A}_2$=diag($b_1,\dots,b_{n-m}$) with the additional property of 
$\l_1\geq\mbox{max}\{|a_1|,\dots,|a_{n-m}|\}$ and its counterpart for 
$\l_2$, ${\bf A}_2$, 
which are the requirements for $\T_1$ and $\T_2$ to be in $\DP^+_2|_x$ (see 
proposition \ref{DP-condition}). Now  
we must find a positive parameter $\alpha$ such that 
$\T_1-\alpha\T_2$ is in $\DP^+_2|_x$ which will be
achieved if we take $\alpha$ fulfilling the inequalities
\be
\l_1-\alpha\l_2\geq|a_{j}-\alpha b_j|,\ \ \,
\ j=1,\dots,n-m.
\label{inequality}
\ee 
(\ref{inequality}) can
be rewritten as:
\be
\fl
\alpha\l_2-\l_1\leq a_j-\alpha b_j\leq\l_1-\alpha\l_2
\Rightarrow
\left\{\begin{array}{c}
\alpha(\l_2+b_j)\leq\l_1+a_j\\
\alpha(\l_2-b_j)\leq\l_1-a_j
\end{array}\right..
\label{ineq}
\ee
Here $\l_1+a_j> 0 < \l_2+b_j$ (if they were zero, then
dim($Span\{\mu(\T_1)\}$) would be greater than $m$), and   
$\l_1-a_j=0\Leftrightarrow \l_2-b_j=0$ due to the property $\s(\T_1)=\s(\T_2)$
(to see this just compute $\mu(\T_1\times\T_1)$ and $\mu(\T_2\times\T_2)$ 
in the above defined orthonormal basis) so we can always choose a 
positive $\alpha$ fulfilling both inequalities (\ref{ineq}).  

Suppose now that we are in the case with no principal directions. This means  
under our hypotheses that $\s(\T_1)=\s(\T_2)=\emptyset$ from what we conclude 
that the functions $f_1(\k_1,\k_2)=\T_1(\k_1,\k_2)$ and 
$f_2(\k_1,\k_2)=\T_2(\k_1,\k_2)$ are nowhere vanishing on 
$\d\Z^+|_x\times\d\Z^+|_x$. Define the set ${\cal S}^+\subset\d\Z_x^+$ as 
follows (see \cite{RINDLER}): 
choose Minkowskian coordinates $\{x^0,\dots,x^{n-1}\}$ on the vector space 
$T_x(V)$ and let
$\k\in\d\Z^+_x$ be in ${\cal S}^+$ if $x^0=1$ when expressed in these coordinates.
This implies that for such $\k$ its coordinates fulfill the constraint
 $1=(x^1)^2+\dots+(x^{n-1})^2$ and that every null vector $\n\in\d\Z^+_x$ 
can be written as $\n=\beta(\n)\k$ for some $\k\in{\cal S}^+$, $\beta(\n)>0$. 
Hence ${\cal S}^+$ is a compact set of $T_x(V)$ 
which in turn means that it must also be compact in $\Z^+_x$ since 
$\overline{\Z^+_x}=\Z^+_x$. Therefore the continuous functions $f_1$ and $f_2$ 
achieve positive upper and lower bounds on ${\cal S}^+\times{\cal S}^+$,
termed as $M_1$ and $m_1$, $M_2$ and $m_2$, respectively. 
According to this, the inequality 
$f_1(\k_1,\k_2)-\alpha f_2(\k_1,\k_2)\geq m_1-\alpha M_2$ holding for every 
$\k_1,\k_2\in{\cal S}^+$ tells us that we can choose 
$0<\alpha\leq m_1/M_2$ such that $\T_1(\k_1,\k_2)-\alpha\T_2(\k_1,\k_2)\geq 0$ 
from what we deduce that this will also hold 
for every pair of null vectors in $\Z^+_x$ as is clear by just 
rewriting them in terms of their ${\cal S}^+$ partners. This
proves that $\T_1-\alpha\T_2\in\DP^+_2|_x$ for such values of $\alpha$ 
as required.\N

\medskip
\noindent
{\bf Remark.} Note that this proof can be repeated interchanging the roles of 
$\T_1$ and $\T_2$ and so we deduce that there exist a positive constant $\beta$
such that $\T_2-\beta\T_1\in\DP^+_2|_x$.

Our last algebraic result relates the strict {\it null convergence condition} 
for a rank-2 symmetric tensor with the dominant energy condition under certain 
assumptions (recall that the tensor $\T\in T^0_2(x)$ 
fulfills the null convergence condition if $\T(\k,\k)\geq 0$ for 
every null vector $\k$).
\begin{lem}
$\T\in T^0_2(x)$ satisfies $\T(\k,\k)> 0$ $\forall 
\k\in\partial\Z_{x}$ if and only if there is a positive $b$ such that 
$\T_{sym}+b\G|_x\in\DP^+_2|_x$ where $\T_{sym}$ is the symmetric 
part of $\T$. 
\label{convergencenull}
\end{lem}
\P The implication from right to left is obvious so we will assume that 
$\T_{sym}$ satisfies the strict null convergence condition but is not a future 
tensor (otherwise the result is obvious as well). We must prove that, for every
null future-directed vector $k^a$, $(T_{(ac)}+b\rmg_{ac})k^c$ is future-directed,
or equivalently that $(\T_{sym}\times\T_{sym})(\k,\k)+2b\T(\k,\k)\geq 0$ 
$\forall \k\in\partial\Z^+_{x}$. 
Consider the {\em compact} set ${\cal S}^+$ as in lemma \ref{basic-lemma} 
and let $m_1$ and $m_2>0$ be the minima that $\T_{sym}\times\T_{sym}$ and $\T$ 
reach over ${\cal S}^+\times{\cal S}^+$, respectively. Then
$(\T_{sym}\times\T_{sym})(\k,\k)+2b\T(\k,\k)\geq m_{1}+2bm_{2}$ so that it is 
enough to choose $2b\geq -m_{1}/m_{2}$. 
Having proven the result on ${\cal S}^+\times{\cal S}^+$ the 
assertion follows for every null future-directed vector $\k$ by the same 
procedure as in lemma \ref{basic-lemma}.\N

\medskip
\noindent
{\bf Remark.} When $\T$ satisfies the null convergence 
condition and there are some null $\k$ such that $\T(\k,\k)=0$ the 
result also holds requiring that $\T_{sym}(\k,\, )$ be causal for 
these $\k$, which is equivalent to demanding that $\k$ is 
a null eigenvector of $\T_{sym}$.

In this work we deal with smooth sections of the set $\DP^+_r(V)$ and 
so we cannot expect that these sections keep the same algebraic character
at all points of the manifold $V$.
This becomes apparent if we work with rank-2 tensors
so that one can use the tools already developed for the classification of other
well known rank-2 tensors in General Relativity when they are seen as smooth
sections of $T_2(V)$ \cite{RAUL,HALL}. We specially quote the results of 
\cite{RAUL} which fit perfectly well in our scheme. 
Basically the authors show that one can decompose the manifold as the union of 
disjoint open sets (or sets with non empty interior) where 
any symmetric tensor has a constant Segre type plus a set with empty interior 
(see main text
for a further explanation of this).

\subsection{Causal relationship.}
Let us consider two diffeomorphic $n$-dimensional Lorentzian 
manifolds $(V,\G)$ and $(W,\tT)$. We will next 
introduce the following notions and properties taken from PII.
\begin{defi}
A diffeomorphism $\f:V\rightarrow W$ is said to be a causal relation from
$V$ to $W$, denoted as $V\prec_{\f}W$, if $\f'\u\in\Z^+(W)$ for every 
$\u\in\Z^+(V)$. The Lorentzian
manifold $W$ is causally related with $V$, which is denoted simply as $V\prec W$,  
if such a diffeomorphism $\f$ exists. 
\label{causal-rel}
\end{defi}
The most simple example of causal relation is a conformal mapping 
from a Lorentzian manifold to another. Here we remove the null cone 
preservation carried out by conformal transformations and just ask for 
the mapping of the whole Lorentzian cone $\Z^+_x$ into $\Z^+_{(\f(x))}$
for each point of the domain manifold $V$.
It is also possible to define in the same fashion diffeomorphisms which map
$\Z^+_x$ into $\Z^{-}_{\f(x)}$ ({\it anticausal relations}). They can be brought
into causal relations just by choosing the opposite causal orientation for 
the target manifold so we will only consider causal relations in 
this work. 

Causal relations have the following interesting properties
shown in PII:                                                                      
\begin{prop}
The following statements are equivalent for two diffeomorphic Lorentzian 
manifolds $(V,\G)$ and $(W,\tT)$.
\begin{enumerate}
\item $V\prec_{\f}W$
\item $\f^*(\DP^+_r(W))\subseteq\DP^+_r(V)$ $\forall r\in\NAT$
\item $\f^*(\DP^+_r(W))\subseteq\DP^+_r(V)$ for a given odd $r\in\NAT$.
\end{enumerate}
In addition to this $\prec$ is a transitive relation i.e. $U\prec V$ and 
$V\prec W$ implies $U\prec W$. Nonetheless it is not an antisymmetric relation.\N
\label{basic-prop}
\end{prop}
Although this proposition states important properties of causal relations it 
fails to give a practical procedure to find out if a given diffeomorphism is
a causal relation. Next theorem from PII fills up this gap:
\begin{theo}
A diffeomorphism $\f:V\rightarrow W$ satisfies $\f^{*}\tT\in\DP^+_2(V)$ if and 
only if $\f$ is either a causal or an anticausal relation.\N
\label{basic-prop2}
\end{theo}
This result is quite valuable in the study of causal relations and
causal symmetries because we can use the results dealing with causal
tensors to study and classify these transformations.  Nonetheless it
is also possible to give another equivalent condition upon the tensor
$\f^*\G$ which will turn out to be very useful as well.

\begin{theo}
The diffeomorphism $\f:V\rightarrow W$ is either a causal or an anticausal 
relation iff $\f^*\tT|_x(\k,\k)\geq 0$, $\forall\ \k\in \partial\Z_{x}$ 
and for every $x\in V$.
\label{nullconvergence}
\end{theo}
\P One implication is obvious, so assume that 
$0\leq \f^*\tT|_x(\k,\k)=\tT|_{\f(x)}(\f'\k,\f'\k)$ for all null $\k$. 
This tells us that $\f'\k$ is causal for every null $\k$ so we must only show 
that the causal orientation is consistently maintained for all them.
Pick a pair of null vectors $\k_1$, $\k_2\in\Z^+_x(V)$ such that 
$\tT|_{\f(x)}(\f'\k_1,\f'\k_1)>0$ (if there is no such $\k_{1}$ then every null 
vector goes to a null vector so that $\f$ is actually a conformal relation,
see theorem \ref{CONF} below) and assume that $\f'\k_1\in\Z^+_{\f(x)}(W)$, 
$\f'\k_2\in\Z^-_{\f(x)}(W)$. This would mean $\tT|_{\f(x)}(\f'\k_1,\f'\k_2)<0$ 
so if we construct a continuous family of nonvanishing null vectors 
$\n_{\l}\in\Z^+_x(V)$, $\l\in[0,1]$ with the properties $\n_0=\k_1$, $\n_1=\k_2$ 
and $\n_{\l_1}\neq\n_{\l_2}$ $\forall\ \l_1\neq\l_2$ the function 
$f(\l)=\tT|_{\f(x)}(\f'\k_1,\f'\n_{\l})$ should vanish at a certain 
$\bar{\l}\in (0,1)$. This would imply that  
$\f'\k_1$ and $\f'\n_{\bar{\l}}$ are null and proportional to each 
other, which is impossible as $\k_1$ and $\n_{\bar{\l}}$ are not proportional 
by hypothesis.\N

Since $\prec$ is not a symmetric relation we conclude that it is a 
{\it preorder} in the set of Lorentzian manifolds. We may also
have the possibility of $W$ being causally related with $V$ but not the other
way round (denoted as $W\not\prec V$) which happens for instance if certain
global causal properties holding for one of the manifolds does not for 
the other. This leads to
\begin{defi}
Two diffeomorphic Lorentzian manifolds $(V,\G)$ and $(W,\tT)$ are called 
isocausal, written as $V\sim W$, if $V\prec W$ and $W\prec V$. 
\label{isocaus-def}
\end{defi} 
According to this definition we can gather Lorentzian manifolds in 
equivalence classes defined by the equivalence relation $V\sim W$ 
$\Leftrightarrow$ $V\prec W$ and $W\prec V$. Each equivalence class 
 coset$(V)$ can be thought of as a causal structure on 
the underlying manifold $M$. Indeed it 
is possible to define a partial order $\preceq$ 
in Lor($M$)/$\sim$ by 
$$
\mbox{coset}(V)\preceq \mbox{coset}(W)\Leftrightarrow V\prec W,
$$ 
thus generalizing the well known
hierarchy of causality conditions used in General Relativity.
The relevance of $\sim$ and $\preceq$ in causality theory was addressed in PII.
                               
A first classification of the causal relations can be
performed according to the set of null vectors which
remain null under its push-forward at a fixed point $x$ (called {\it canonical
null directions} of the transformation at $x$). This set, denoted
as $\mu(\f)|_x$ for each causal relation $\f$, constitutes the part of the 
null cone preserved by $\f$ at $x$ so we get a gradation starting at the case 
 where the full null cone is preserved and ending in the case without
canonical null directions with a range of situations lying in between.
The next result (see PII) permits us to calculate $\mu(\f)|_x$.  
\begin{prop}
The set $\mu(\f)|_x$ of canonical null directions of a causal relation 
$\f:V\rightarrow W$ at $x$ is given by $\mu(\f^{*}\tT|_x)$.
\label{MUFI}
\end{prop}
\P The equality $\tT|_{\f(x)}(\f'\k,\f'\k)=\f^{*}\tT|_x(\k,\k)$ implies that 
if a given null vector $\k\in\Z^+_x$ goes to a null vector $\f'\k$ 
then $\k\in\mu(\f^{*}\tT|_x)$ 
which proves the inclusion
$\mu(\f)|_x\subseteq\mu(\f^{*}\tT|_x)$. The same equality 
serves also to prove 
the other inclusion.\N

With this result at hand we conclude that the same remarks made for the set 
of principal directions of a causal tensor hold for the canonical null
directions of a causal relation at a point $x$. These local considerations can 
also be turned into global ones if we restrict our study to regions 
of the manifold in which the algebraic type of $\f^{*}\tT$ does not change.
Therefore, when we speak about the canonical null directions of a causal relation 
we will implicitly assume that the algebraic type of $\f^{*}\tT$ does not change 
over the manifold for the transformation $\f$. Another important point is that we do
not need to take $\s(\f^*\tT|_x)$ into account.
\begin{prop}
$\s(\f^*\tT|_x)=\mu(\f^*\tT|_x)$ for a causal relation $\f$.
\label{NULL-PRINCIPAL}
\end{prop}
\P We already know the inclusion $\mu(\f^*\tT|_x)\subseteq\s(\f^*\tT|_x)$. If the
couple of null vectors $\k_1$ and $\k_2$ are in $\s(\f^*\tT|_x)$ then 
$0=(\f^*\tT)|_x(\k_1,\k_2)=\tT|_{\f(x)}(\f'\k_1,\f'\k_2)$ which is only possible 
if $\f'\k_1$ and $\f'\k_2$, being both future-directed, are null and 
proportional meaning that $\k_1$ (and $\k_2$) are in $\mu(\f^*\tT|_x)$ 
which leads to the other inclusion.\N

The transitivity of the causal relation $\prec$ is due to the 
fact that the composition $\f_2\circ\f_1$
of two causal relations $\f_1:U\rightarrow V$ and $\f_2:V\rightarrow W$ is 
also a causal relation. Nonetheless the inverse of a causal relation will in
general fail to be a causal relation except in a very special case
(see PII for the proof).
\begin{theo}
For a diffeomorphism $\f : V \longrightarrow W$ the following
properties
are equivalent:
\begin{enumerate}
\item $\f$ is a causal (or anticausal) relation with $n$ linearly independent 
canonical null directions (i.e. every null vector goes to a null vector).
\item $\f^* \tT =\lambda \G$, $\lambda >0$.
\item $(\f^{-1})^* \G =\mu \tT$, $\mu>0$.
\item $\f$ and $\f^{-1}$ are both causal (or anticausal)
relations.
\end{enumerate}\N
\label{CONF}
\end{theo}
Therefore we see that a conformal relation $\f:V\rightarrow W$ is 
characterized as a diffeomorphism such that $V\prec_{\f} W$ and 
$W\prec_{\f^{-1}}V$ from what we deduce that the only groups of causal 
relations are formed exclusively by conformal diffeomorphisms. 

  Another result of PII needed in this paper deals with how the causal 
character of causal vectors changes under a causal relation.
\begin{prop}
If $V\prec_{\f}W$ then:
\begin{enumerate}
\item $\u\in\Z^{+}(V)$ is timelike $\Longrightarrow\,\, \f'\u\in
\Z^{+}(W)$
is timelike.
\item $\u\in\Z^{+}(V)$ and $\f'\u\in \Z^{+}(W)$ is null
$\Longrightarrow \u$ is null.
\item $\K\in\DP^+_1(W)$ is timelike $\Longrightarrow\,\,
\f^{*}\K\in \DP^{+}_1(V)$ is timelike.
\item $\K\in\DP^+_1(W)$ and $\f^{*}\K\in
\DP^{+}_1(V)$
is null $\Longrightarrow \K$ is null.
\end{enumerate}\N
\label{caus-trans}
\end{prop}    

\section*{Appendix B}
\setcounter{section}{2}
\setcounter{theo}{0}
 In this appendix we prove that every linear automorphism 
$\hat{\T}\in\C(\L,\bfeta)$ has at least 
an invariant causal direction, i.e there exists an element $\u_x\in\Z^+_x$ 
with the property $\hat{\T}\u_x\propto\u_x$. This result may well
be available in other places of the literature under a different terminology 
although we have preferred to a proof adapted to our work 
(see for instance theorem 1.10.3 of \cite{BAPAT} for an indirect proof 
in a special case). To start with, a preliminary result is needed \cite{BROUWER}.
\begin{theo}[Brouwer's fixed point theorem]
Define $B^n=\{x\in\r^n:|x|\leq 1\}$ and let $\f:B^n\rightarrow B^n$ be a continuous
map. Then there exists a point $x_0\in B^n$ such that $\f(x_0)=x_0$. 
\label{fixedpoint}
\end{theo}  
\begin{theo}
For every linear $\hat{\T}\in\C(\L,\bfeta)$ we can always find a vector 
$\u\in\Z^+_x$ such that $\hat{\T}(\u)=\beta\u$, $\beta>0$.
\label{fixedvector}
\end{theo}
\P We start our proof by setting an equivalence relation in the vector space $\L$ 
by means of the 
definition $\u_1\sim\u_2\Leftrightarrow\u_1=b\u_2$ for some $b\in\r$. The elements of 
$\L/\sim$
will be denoted with a square bracket enclosing a representative $\u$ of each class as $[\u]$.
Clearly, it makes sense the definition $\hat{\Z}_x=\{[\u]\in \L:\u\in\Z_x\}$ as 
$b\u\in\Z_x$ $\forall b\in\r$ if $\u\in\Z_x$. Now, define a system of Minkowskian coordinates 
$\{x^0,x^1,\dots x^{n-1}\}$ in 
$(\L,\bfeta)$ and write every vector $\u\in \L$ in these coordinates as 
$\u=(a,ay^1,\dots,ay^{n-1})$, $a\in\r$. If $\u\in\Z_x$ then $(y^1)^2+\dots+(y^{n-1})^2\leq 1$
so it is possible to define a map $\hat{\Psi}_X:\hat{\Z}_x\rightarrow B^{n-1}$ by
$\hat{\Psi}_X([\u])\equiv(y^1,\dots,y^{n-1})$ where the subscript $X$ means that this map
depends on the chosen Minkowskian coordinate system. $\hat{\Psi}_X$ is a well-defined 
one-to-one map between $\hat{\Z}_x$ and $B^{n-1}$ which means that 
$\hat{\Psi}^{-1}_X:B^{n-1}\rightarrow\hat{\Z}_x$ exists and is also one-to-one.
Any $\hat{\T}\in\C(\L,\bfeta)$ is characterized by the properties of being linear and 
$\hat{\T}(\Z^+_x)\subseteq\Z^+_x$ so $\hat{\T}(\Z_x)\subseteq\Z_x$ and we can define the map
$\hat{\f}:\hat{\Z}_x\rightarrow\hat{\Z}_x$ by $\hat{\f}([\u])\equiv[\hat{\T}(\u)]$, $\forall\u\in\Z_x$.
Again this is a well-defined map and from it we construct another one given by 
$\hat{\Psi}_X\hat{\f}\hat{\Psi}^{-1}_X:B^{n-1}\rightarrow B^{n-1}$ which is continuous because it is the
composition of continuous maps. According to Brouwer fixed point theorem, we know that there must exist
a point $p\in B^{n-1}$ such that $\hat{\Psi}_X\hat{\f}\hat{\Psi}^{-1}_X(p)=p\Rightarrow
\hat{\f}\hat{\Psi}^{-1}_X(p)=\hat{\Psi}^{-1}_X(p)$ which means that 
$[\u_p]\equiv\hat{\Psi}^{-1}_X(p)\in\hat{\Z}_x$ is a fixed point of $\hat{\f}$ as well. Therefore
$\hat{\f}[\u_p]=[\hat{\T}(\u_p)]=[\u_p]\Rightarrow\hat{\T}(\u_p)=a_p\u_p$ with $\u_p\in\Z_x$ and $a_p\in\r$.
Given that the causal orientation of $\u_p$ must be preserved under $\hat{\T}$ 
we conclude that $a_p$ is a positive constant thus finishing the proof. \N

\end{document}